\documentclass[%
reprint,
amsmath,amssymb,
pra,
]{revtex4-2}

\usepackage{graphicx}
\usepackage{dcolumn}
\usepackage{bm}
\usepackage{capt-of}
\usepackage{wrapfig}

\usepackage{blindtext}
\usepackage{graphicx, caption}
\usepackage{multirow}
\usepackage{float}
\usepackage{subfig}
\usepackage{geometry}
\usepackage{amsmath,amssymb}
\usepackage{amsthm}
\usepackage{bbold}
\usepackage{mathtools}
\usepackage{braket}
\usepackage{booktabs}
\usepackage{hyperref}
\hypersetup{colorlinks=true,linkcolor=blue}

\begin{document}
	\title{{{Entanglement Degradation in the Presence of Markovian Noise: a Statistical Analysis}}}
	\author{Nunzia Cerrato}
	\affiliation{Scuola Normale Superiore, Piazza dei Cavalieri 7, I-56126 Pisa, Italy}
	\author{Giacomo De Palma}
	\affiliation{Department of Mathematics, University of Bologna, 40126 Bologna, Italy}
	\author{Vittorio Giovannetti}
	\affiliation{NEST, Scuola Normale Superiore and Istituto Nanoscienze-CNR, Piazza dei Cavalieri 7, I-56126 Pisa, Italy}
	
	\begin{abstract}

{{
Adopting a statistical approach we study the degradation of entanglement of a quantum system under the action of an ensemble of randomly distributed Markovian noise. This enables us to address  scenarios where only limited information is available on the mechanisms that rule the noisy evolution of the model. As an application, we characterize the statistic of entanglement deterioration for a quantum memory formed by $n$ qudits that undergo randomly distributed local, uniform, Markovian noise evolution. 
}}
\end{abstract}
	
	\maketitle
	
	\section{Introduction}
	Entanglement represents a unique yet fragile feature of quantum mechanics: it can be used as a resource in famous quantum information protocols, like superdense coding \cite{Quantum_dense_coding} and quantum teleportation \cite{Teleportation,Ent_swapping}, as well as in contexts like quantum computation and quantum cryptography \cite{Horodecki}, but, at the same time, it can be easily degraded by noise, making difficult to exploit all its potential.
	For this reason, it is meaningful to study and deepen the entanglement properties of quantum states in the framework of open quantum systems, which are systems whose dynamics is influenced by the interaction with noisy external environments.
	The first works in this field \cite{Ent_dynamics_1,Ent_dynamics_2} showed that the results of such an interaction are decoherence of individual systems and entanglement degradation, which are two different aspects of the same process. However, even if coherence can vanish asymptotically in time, a different effect has been discovered regarding entanglement: a total loss of entanglement could occur, after a finite time, even in presence of weakly dissipative environments. This effect has been called {{Entanglement Sudden Death}} and it has been the main object of recent studies, which addressed this phenomenon both on a theoretical level
	and, subsequently, on an experimental one. An in-depth treatment of entanglement dynamics for open quantum systems, especially focused on decoherence effects, can be found, among others, in the work of Aolita et al. \cite{OQSD}, where the Authors give a throughout review of this phenomenon both considering bipartite and multipartite systems, while also taking into account experimental results in this field.
				\begin{figure}
		{\includegraphics[scale=0.5]{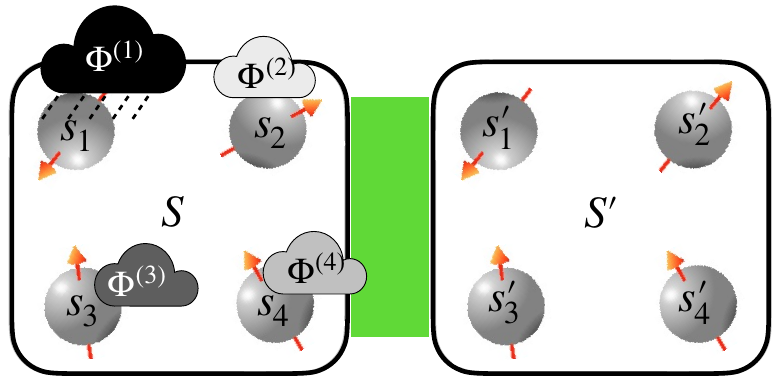}}
		\caption{{{Pictorial representation of the degradation of the entanglement 
		between two quantum memories $S$, $S'$  formed by large collection of $n$ qudits, induced by random noise
		that acts locally, and independently on the subparts  $s_1,s_2, \cdots, s_n$  of $S$. }}}\label{figmany1}
	\end{figure}
	What most studies on this subject have in common is that they consider well-characterized noises. However, in real contexts, noise is not necessarily known or, in other words, it is not necessarily said that we know exactly the mathematical structure of noise and how quantum systems are expected to evolve when subjected to it. A practical example is given by quantum computers, where one can have a sudden and unexpected change in voltage or temperature, or the appearance of electromagnetic fields due to nearby quantum objects, effectively causing total randomness in the possible noises that could afflict a qubit. Very often, all that is known is that the different types of noise acting on a system share only some basic characteristics, for example they can be of the Markovian type, and it would be desirable to exploit this information to obtain a general understanding about how systems behave in these cases, with respect to the physical properties of interest.

	{{In this spirit, we have decided to  study the entanglement degradation that  a quantum system $S$ has initially
	established with an auxiliary  reference $S'$ 
	 under the influence of an ensemble of (local) Markovian noises.}} 
The starting point of our
	analysis is the distribution of what we call the
	 positive partial transpose time (PPTT) of the model. This quantity
	represents the minimum time after
	which the partial transponse of a maximally entangled state between $S$ and $S'$ 
	becomes positive, hence not-distillable. An analogous quantity called \textit{entanglement survival time} (EST)
	was introduced in~\cite{Gatto}: this last measures the minimal time at which the input entanglement of the model
	gets completely degraded. For each individual noisy Markovian evolution, the two terms are connected by a natural ordering which makes the PPTT a lower bound  for the EST. The main advantage of focusing on PPTT instead of EST is associated with the fact that the former is relatively easy to compute (at least for low-dimensional systems), while the latter is typically a difficult functional to compute. 
A case of particular interest is obtained when  $S$ corresponds to a quantum memory formed by 
$n$-qudits $s_1,\cdots, s_{n}$ that evolve under the action of random Markovian noisy evolutions that act locally and independently on each individual subsystem (see Fig.~\ref{figmany1}). For these models, the PPTT distribution turns out 
to be a 
simple functional of the PPTT distributions of the individual subsystems of $S$, leading to some important
simplification.

As an application of our analysis we focus on the case where $S$ (or its subsystems $s_i$) corresponds either to  a qubit or to a qutrit evolving under the action of a dynamical semigroup whose Lindbladian generator is 
selected via the random matrix sampling method developed in Ref.~\cite{Zyczkowski}. 
Keeping fixed the ratio between the Hamiltonian and dissipative contributions of the generators, 
the resulting PPPT distributions are studied with the help of a statistical analysis, carried out with the Minitab software~\cite{Minitab}. We also exhibit   probability density functions that represent a good fit for the recovered data and we investigate the relationship between their characteristic times (such as mean time, median time, and minimum time) and the model parameters. Finally, since we report a limit trend in these distributions, appearing for predominant unitary Hamiltonian terms in the dynamical generator, we propose a general analytical expression for the dynamical generator in this limit.

	 {{The present work is organized as follows. Sec.~\ref{sect:EST} introduces the notation and the
	key definitions. In Sec.~\ref{sec:local} we formally define the PPTT distribution, its cumulative counterpart, and clarify how they behave when applied to many-body quantum systems that evolve under the action of independent local noise models.  
	In Sec.~\ref{sec:samp} we specify the problem to the case of qubit and qutrit systems
	subjected to random dynamical semigroups sampled with the measure introduced in Ref.~\cite{Zyczkowski}.
	The obtained numerical  results are reported in Sec.~\ref{sect:ESTs_distrib}.   In particular in Sec.~\ref{sect:LimLind} we focus on the limit trend emerging  in the  PPTT distribution when considering predominant unitary Hamiltonian contributions, proposing a general analytical expression of the dynamical generator that, properly sampled, allows one to obtain the limit distribution. Section~\ref{sec:stats} is
	devoted to provide a statistical analysis of the results of Sec.~\ref{sect:ESTs_distrib}.
	Conclusions and final remarks are presented in Sec.~\ref{sect:conclusion}.
	The manuscript includes also a couple of technical appendixes. }}
{{	
	\section{Entanglement degradation for dynamical semigroups}\label{sect:EST}}}

 	The temporal evolution of
	a quantum {{system $S$}} that is interacting with an external environment 
	can be formally  represented in terms  of  a
	one-parameter family $\mathcal{F}:=\{\Phi_{(0,t)},t\ge0\}$ of linear completely positive and trace-preserving  (LCPT) superoperators $\Phi_{(0,t)}$~\cite{Petruccione}.
	In this picture, given {{$\hat{\rho}_S(0)$}} the density matrix 
	that defines the input {{state  of $S$}}	
	at time $t=0$,
	its output counterpart at time $t\geq 0$ is obtained via the  application of 
	$t$-th element of the set via the mapping 
	{{\begin{eqnarray} 
	\hat{\rho}_S(0) \mapsto \hat{\rho}_S(t)= \Phi_{(0,t)}(\hat{\rho}_S(0))\;. \label{def1} 
	\end{eqnarray} }}		
	 In our analysis we shall focus on dynamical processes that are  of Markovian type and
	 homogeneous with respect to $t$.
	  Under these conditions the family ${\cal F}$ forms a so-called dynamical semigroup, 
characterized by a hierarchical 
	   ordering of the  elements $\Phi_{(0,t)}$ which, as time increases, leads to a gradual accumulation of the deterioration effects induced by the environment.  The corresponding evolution is described  by a  Gorini, Kossakowski, Sudarshan, and Lindblad (GKSL) master equation,
	   {{
	   \begin{eqnarray} \label{dynsemigroup} 
	   \dot{\hat{\rho}}_S(t) = {\cal L}(\hat{\rho}_S(t))\; \Longrightarrow\;  \Phi_{(0,t)} = e^{{\cal L} t} \;, \end{eqnarray} }}
	 whose  dynamical generator
 $\mathcal{L}$, known as Lindblad superoperator or {\it Lindbladian}~\cite{LGKS}, completely characterizes the properties of ${\cal F}$. 
Such a term can	 be expressed as the sum of two components: a unitary Hamiltonian term, represented by a self-adjoint operator $\hat{H}$, which preserves quantum coherence of {{$S$}}, and a purely dissipative term which instead tends to suppress coherence. 
Specifically, given $\{\hat{F}_{m}\}_{m=1,N^2 -1}$ a set of traceless matrices that satisfy the orthonormality condition $\text{Tr}[\hat{F}_{m}^{\dagger}\hat{F}_{n}]=\delta_{mn}$, with $N$ being the dimension of the Hilbert space {{${\cal H}_S$  of $S$}}, the  Lindbladian of a generic  dynamical semigroup ${\cal F}$ can be written as
	{{\begin{equation}\label{eq:Lindblad_ME}
		\mathcal{L}(\cdots) = -i[\hat{H},\cdots] +  {\cal D}_K(\cdots) \;,
	\end{equation}}}
	with the dissipative contribution given by 
		\begin{eqnarray}\label{eq:Dissipator_Kmn}
			 {\cal D}_K(\,\cdots\,)&=& \sum_{m,n=1}^{N^2 -1}K_{mn}\Bigg(\hat{F}_{n}(\,\cdots\,)\hat{F}_{m}^{\dagger}\\&&-\frac{1}{2}\Big( \hat{F}_{m}^{\dagger}\hat{F}_{n}(\,\cdots\,) + (\,\cdots\,)\hat{F}_{m}^{\dagger}\hat{F}_{n} \Big)\Bigg), \nonumber 
	\end{eqnarray}
where 
 $K_{mn}$ are elements of a positive definite matrix $K$, called Kossakowski matrix 
 {{(hereafter for simplicity $\hbar$ is set equal to 1)}}. 
For future reference we remind that upon diagonalization of $K$, (\ref{eq:Dissipator_Kmn}) can also be casted in the more compact form 
		\begin{eqnarray}\label{eq:Dissipator_Kmn_lind}
			 &&\!\! \!\! \!\! \!\!{\cal D}_K(\,\cdots\,)= \sum_{n=1}^{N^2 -1}\Bigg(\hat{L}^{(n)}_K(\,\cdots\,)\hat{L}^{(n)\dagger}_K\\\qquad &&-\frac{1}{2}\Big( \hat{L}^{(n)\dagger}_K\hat{L}^{(n)}_K(\,\cdots\,) + (\,\cdots\,)\hat{L}^{(n)\dagger}_K\hat{L}^{(n)}_K\Big)\Bigg), \nonumber 
	\end{eqnarray}
with  $\hat{L}^{(n)}_K := \sqrt{\lambda^{{(n)}}_K} \sum_{m}   U^{(m,n)}_K \hat{F}_{m}$ the associated Lindblad operators constructed in terms of 
 the eigenvalues $\lambda^{{(n)}}_K$ of $K$ and of matrix elements  $U^{(m,n)}_K$ which define  its eigenvectors through the spectral decomposition 
$K_{n,m} = \sum_{n'} U^{(n,n')}_K U^{(m,n')*}_K  \lambda^{{(n')}}_K$.

	\subsection{Entanglement Survival Time and Positive Partial Transpose Time}

 Positive Partial Transpose (PPT) channels  and 
 	Entanglement Breaking (EB) channels are LCPT maps that represent the most severe examples of noisy transformations  a quantum system can experience. 
	When PPT channels are applied locally to the joint input states $\hat{\rho}_{SS'}$  of {{system $S$}} and an external auxiliary system $S'$, the resulting joint outputs have a positive partial transpose, indicating that they are not entanglement-distillable. In contrast, when EB channels are used, the resulting density matrix is separable making any EB channel is also PPT (the opposite being
	not necessarily true for $N\geq 3$)~\cite{PPT}. 
	
The initial elements  of  a dynamical semigroup ${\cal F}$	are usually close to the identity mapping and do not exhibit the properties of being either EB or PPT. However, as time progresses and noise accumulates, it is highly likely that the maps $\Phi_{(0,t)}$  will acquire these characteristics. In order to assess the influence of the dynamics ${\cal F}$ on {{system $S$}}, it is therefore sensible to define two characteristic times~\cite{Gatto}: the  {\it entanglement survival time} (EST)  and the {\it positive partial transpose time} (PPTT)
 defined respectively as	
	\begin{eqnarray}
		\tau_{est}({\cal L}) &:=& \min\{ t\ge0 \text{ s.t. } \Phi_{(0,t)} \in \text{EB}\}, \\ 
		\tau_{ppt}({\cal L}) &:=& \min\{ t\ge0 \text{ s.t. } \Phi_{(0,t)} \in \text{PPT}\}, \label{ppttime} 
	\end{eqnarray} 
	where the notation stresses that ESTs and PPTTs  are explicit functionals of the generator~${\cal L}$ of the  semigroup
	(when no finite $t$  exists such that $\Phi_{(0,t)} \in \text{EB}$  we simply set 
	$\tau_{est}({\cal L})=\infty$, similarly for $\tau_{ppt}({\cal L})$). 
	{{It is worth stressing that since in our analysis the $\Phi_{(0,t)}$'s are assumed to be elements of a  dynamical semigroup~(\ref{dynsemigroup}),
	it naturally follows that once the dynamics acquires an EB or PPT property, it will keep forever, i.e. 
	\begin{eqnarray} 
	&&\Phi_{(0,t)} \in  \text{EB}\;, \qquad\;\;  \forall t\geq \tau_{est}({\cal L}) \;,\label{impo111}  \\
	&&\Phi_{(0,t)} \in  \text{PPT}\;,\qquad  \forall t\geq \tau_{ppt}({\cal L}) \;.  \label{impo222} 
	\end{eqnarray} }}
In case $A$ is a qubit $\tau_{ppt}({\cal L})$  and $\tau_{est}({\cal L})$ coincides: for larger systems however the former will typically represent a lower bound for the latter, i.e. 
\begin{equation}  
\tau_{ppt}({\cal L}) \leq \tau_{est}({\cal L})\;. \end{equation} 
{{From~\cite{EBC} we known  that determining if a quantum map  is EB  is equivalent to check if the associated Choi-Jamiołkowski state~\cite{Choi-Jam}  is separable or not.  
A similar property holds also for the  PPT property (see Appendix~\ref{appePPT} for a proof of this fact).
}}
Accordingly, 
computing $\tau_{est}({\cal L})$ (resp. $\tau_{ppt}({\cal L})$)   corresponds to finding the minimum value of $t$ at which the
state 
	{{\begin{equation}\label{choi} 
		\hat{\rho}_{SS'}^{\Phi_{(0,t)}}:=\Phi_{(0,t)}\otimes {\rm{Id}}_{S'}(|{\Psi_{\max}}\rangle_{SS'}\langle {\Psi_{\max}}|),\end{equation} 
	becomes separable (resp. PPT)  (in this expression $S'$ is an auxiliary system of the same dimension $N$ of $S$, $\ket{\Psi_{\max}}_{SS'}= \frac{1}{\sqrt{N}}\sum_{i=1}^{N}\ket{i}_{S}\otimes \ket{i}_{S'}$ is a maximally entangled state of $SS'$, with $\{\ket{i}_{S/S'}\}$ being orthonormal basis, and ${\rm{Id}_{S'}}$ is the identity super-operator on $S'$).
Due to the absence of universal conclusive criteria for identifying the presence of  entanglement in a system, 
 the characterization of the EST is typically very challenging. The same problem does not hold however for the PPTT which instead can be casted in terms of the following optimization problem: 
	\begin{equation}\label{simplifiedppt} 
		\tau_{ppt}({\cal L})=\min\{t\ge0\ \text{s.t.}\ \mathcal{N}(\hat{\rho}_{SS'}^{\Phi_{(0,t)}})=0\}\;, 
	\end{equation}
	where $\mathcal{N}(\hat{\rho}_{SS'}^{\Phi_{(0,t)}})$ is the negativity of entanglement~\cite{NegEnt}. This last  is  an algebraic entanglement measure that can be expressed as
	\begin{equation}
		\mathcal{N}(\hat{\rho}_{SS'})=\frac{1}{2}\sum_{l}(|\lambda_{l}|-\lambda_{l}),
	\end{equation}
	with $\{\lambda_{l}\}$  the eigenvalues of $\hat{\rho}_{SS'}^{T_{S'}}$ and $T_{S'}$ representing the partial transposition w.r.t. to basis $\{\ket{i}_{S'}\}$ of $S'$. }}

Simple scaling arguments can be used to observe that by 
multiplying the generator of the dynamical group by a positive factor $\beta >0$, the EST and PPTT acquire a moltiplicative prefactor $1/\beta$, i.e.  	\begin{equation}{\label{eq:scaling_law_beta}}
		\tau_{\bullet}(\beta \mathcal{L}) = \frac{1}{\beta}\tau_{\bullet}(\mathcal{L})\;,
	\end{equation}
	where hereafter the  symbol “$\bullet$" is used as a placeholder for {\it ppt} and {\it est}. 
	A nontrivial challenge is to comprehend the respective contributions of the Hamiltonian and dissipator components of ${\cal L}$ in determining the values of the ESTs and PPTTs. This inquiry was explored in Ref.~\cite{Gatto} for specific models, such as single qubit and a restricted class of continuous variable systems, where EST and PPT are known to coincide.
In this context, the authors utilized Eq.~(\ref{simplifiedppt}) to examine the functional relationship of $\tau_{est}(\mathcal{L})$ for Lindbladian generators of the form 
\begin{equation}{\label{eq:Lindbladianalphagamma}} 
		\mathcal{L}_{H,K}^{(\alpha,\gamma)}(\,\cdots \,):= -i  \alpha [\hat{H}, \cdots \,] + \gamma \mathcal{D}_{K}(\,\cdots \,)\;,
	\end{equation}
	 where $\hat{H}$ and $K$ were selected to represent specific types of noises (such as dephasing, thermalization, and depolarization). The positive coefficients $\alpha$ and $\gamma$ were introduced as adjustable parameters that determine the relative strengths of the Hamiltonian and dissipator contributions.
Interestingly, the study reveals that in certain cases of the noise models, incorporating a unitary Hamiltonian component into the dissipative dynamics results in a more rapid decay of entanglement. Specifically, the value of $\tau_{est}({\cal L})$ decreases as $\alpha$ increases and approaches a limit as $\alpha$ approaches infinity.
 Our objective is to extend this analysis by adopting a statistical perspective. Instead of focusing on restricted classes of generators, we aim to employ tools from Random Matrix Theory to sample the values of {{$\tau_{ppt}(\mathcal{L})$}} across the entire range of possible ${\cal L}$, following the approach outlined in~\cite{Zyczkowski}. This approach offers two significant advantages: Firstly, it allows us to gain insights into the problem that are not dependent on specific characteristics of the noise model. Secondly, it provides an effective characterization of scenarios where only partial information about the noise affecting the quantum system is available.
{{
\section{PPTT distribution and cumulative PPTT distribution} 
 \label{sec:local} 
 Consider the case where the system $S$ evolves in time under the action of dynamical semigroup 
 $\Phi_{(0,t)}$ whose Lindblad generator  ${\cal L}$ is known  only up to a certain precision. Specifically in our analysis we shall assume that the only prior information we have on the process  is that ${\cal L}$ has been randomly extracted 
  via some  assigned
 problability measure $d\mu({\cal L})$ that acts on the set ${\mathfrak L}$ of the Lindblad superoperators. }}
 Under this condition we define $P_{ppt}(\tau)$ the {{density probability distribution that the selected dynamics 
 has PPTT equal to $\tau$, i.e. formally 
 \begin{equation} \label{mpodef} 
 P_{ppt}(\tau) d\tau : = \mbox{Prob}\Big( \tau_{ppt}({\cal L}) \in[ \tau,\tau+\delta \tau[ \Big)\;. 
 \end{equation} }}
The distribution (\ref{mpodef}) is an emergent property  of the model which can in principle be reconstructed via Eq.~(\ref{simplifiedppt})  by sampling $\mathfrak{L}$ through the measure $d\mu({\cal L})$.
Recalling Eq.~(\ref{impo222}) 
it follows that cumulative integral of $P_{ppt}(\tau)$ yields the probability
$\bar{P}_{ppt}(T)$ that the dynamical  evolution of $S$ at time $T$ is ruled by a PPT map,
\begin{eqnarray} \label{cumu} 
\bar{P}_{ppt}(T) : = \int_0^T d\tau P_{ppt}(\tau)\;. 
\end{eqnarray} 
Operationally  $\bar{P}_{ppt}(T)$ corresponds to the probability that  a maximally entangled state of
  $SS'$,  will become  PPT at time $T$
given the uncertainty on the dynamics of $S$. 
{{Notice that by construction  $\bar{P}_{ppt}(T)$ is null for $T=0$ (the input state being indeed maximally entangled), while in the large $T$ limit,  it   always approaches $1$, {{at least  for those models where  $P_{ppt}(\tau)$ is not zero 
for at some finite $\tau$.}}
By construction $\bar{P}_{ppt}(T)$ also provides a (strict) lower bound, for the probability $\bar{P}_{ppt}(T|\hat{\rho}_{SS'})$
that, starting 
 from a generic (possibly not maximally entangled) input state $\hat{\rho}_{SS'}$ of $SS'$, we find it 
in a PPT configuration at time $T$, i.e.
\begin{eqnarray} 
\bar{P}_{ppt}(T|\hat{\rho}_{SS'}) \geq \bar{P}_{ppt}(T) \;, \qquad \forall \hat{\rho}_{SS'}\;,
\end{eqnarray} 
the inequality being saturated for $\hat{\rho}_{SS'}$ being maximally entangled
(a proof of this follows immediately from the observation in App.~\ref{appePPT}). }}
\subsection{PPTT under Random Local Noise} \label{sec:ppt-local} 
A case of particular interest in quantum computing and information theory is represented by the scenario where 
$S$ is a large quantum memory constituted by a collection of $n$ 
non-interacting quantum systems $s_1,s_2,\cdots, s_n$ which undergo to independent, random, local noisy evolutions as in the scheme depicted in Fig.~\ref{figmany1}.
{{ In this case 
the Lindbladian ${\cal L}$ is given by a sum local terms,
\begin{eqnarray} {\cal L}= {\cal L}^{(1)} + \cdots + {\cal L}^{(n)}\;, 
\end{eqnarray} 
where for $i\in \{ 1,\cdots, n\}$, ${\cal L}^{(i)}$ is a superoperator acting on the $i$-th substems,
the measure $d\mu({\cal L})$   factorizes into product of local measures 
\begin{eqnarray} 
d\mu({\cal L}) = d\mu({\cal L}^{(1)})\cdots d\mu({\cal L}^{(n)})\;,
\end{eqnarray} 
and the
maps $\Phi_{(0,t)}$ reduce to 
\begin{eqnarray} 
\Phi_{(0,t)} = \Phi^{(1)}_{(0,t)}\otimes \cdots\otimes  \Phi^{(n)}_{(0,t)}
\end{eqnarray} 
with  $\Phi_{(0,t)}^{(i)}=e^{{\cal L}^{(i)} t}$.}}
Accordingly, also the  Choi-Jamiołkowski state of the process $\Phi_{(0,t)}$ factorizes 
in the tensor  product of local contributions, 
\begin{eqnarray} \label{decon} 
\hat{\rho}_{SS'}^{\Phi_{(0,t)}}  =\hat{\rho}_{s_1s_1'}^{\Phi_{(0,t)}^{(1)}} \otimes \cdots\otimes  
\hat{\rho}_{s_n s_n'}^{\Phi_{(0,t)}^{(n)}}\;, 
\end{eqnarray} 
with $\hat{\rho}_{s_is_i'}^{\Phi_{(0,t)}^{(i)}}$ 
the Choi-Jamiołkowski  state of ~$\Phi_{(0,t)}^{(i)}$ constructed by applying the latter to a maximally entangled state of
{{$s_i$ with an (isodimensional) component $s_i'$ of $S'$. }}
From Eq.~(\ref{decon}) it now follows that 
 $\hat{\rho}_{SS'}^{\Phi_{(0,t)}}$ is PPT if and only iff all the $\hat{\rho}_{s_is_i'}^{\Phi_{(0,t)}^{(i)}}$ fulfils the same property, i.e. 
\begin{equation} 
\left( \hat{\rho}_{SS'}^{\Phi_{(0,t)}} \right)^{T_{S'}} \geq 0 \quad \Leftrightarrow \quad  \left( \hat{\rho}_{s_is'_i}^{\Phi^{(i)}_{(0,t)}} \right)^{T_{s'_i}} \geq 0 \quad  \forall i \;.
\end{equation} 
Thanks to this implication we can hence conclude that the  cumulative distribution 
$\bar{P}^{[{\rm loc},n]}_{ppt}(T)$ of $S$ can be expressed as the product of the corresponding individual terms, i.e. 
\begin{eqnarray} \label{easyrecursive} 
\bar{P}^{[{\rm loc},n]}_{ppt}(T) =  \bar{P}^{(1)}_{ppt}(T)\cdots  \bar{P}^{(n)}_{ppt}(T)\;,
\end{eqnarray} 
from which one can recover ${P}^{[{\rm loc},n]}_{ppt}(\tau)$ by inversion of~(\ref{cumu}), i.e. 
\begin{eqnarray} \label{easyrecursive1} 
{P}^{[{\rm loc},n]}_{ppt}(\tau) &=& \frac{d  \bar{P}^{[{\rm loc},n]}_{ppt}(\tau)}{d\tau}\\ \nonumber 
&=& \bar{P}^{[{\rm loc},n]}_{ppt}(\tau)  \sum_{i=1}^n   \frac{{P}^{(i)}_{ppt}(\tau)}{\bar{P}^{(i)}_{ppt}(\tau)} \;,
\end{eqnarray} 
(in the above expressions ${P}^{(i)}_{ppt}(\tau)$ and $\bar{P}^{(i)}_{ppt}(T)$ represent respectively the
PPTT distribution of the $i$-th subsystem and its associated cumulative probability).
A simple recursion argument can be used to show that to 
prove Eq.~(\ref{easyrecursive}) it is sufficient to show that such identity holds for $n=2$. {{In this case the probability
${P}^{[{\rm loc},2]}_{ppt}(\tau)$ of selecting a generator ${\cal L}$ of a dynamical process which becomes PPT at time $\tau$}} can be expressed as
the probability of generating a dynamical process  on one of the two subsystems which becomes PPT for the first time exactly at $t=\tau$ times the probability that on the other subsystem we selected a process that becomes PPT at some prior time. In formula this means 
\begin{eqnarray} 
{P}^{[{\rm loc},2]}_{ppt}(\tau) &=& {P}^{(1)}_{ppt}(\tau) \bar{P}^{(2)}_{ppt}(\tau) + \bar{P}^{(1)}_{ppt}(\tau) {P}^{(2)}_{ppt}(\tau)\nonumber \\
&=& \frac{d \bar{P}^{(1)}_{ppt}(\tau)}{d\tau} \bar{P}^{(2)}_{ppt}(\tau) + \bar{P}^{(1)}_{ppt}(\tau)\frac{d \bar{P}^{(2)}_{ppt}(\tau)}{d\tau} \nonumber \\
&=& \frac{d}{d\tau} \left( \bar{P}^{(1)}_{ppt}(\tau) \bar{P}^{(2)}_{ppt}(\tau)\right)\;,\label{ppt222} 
\end{eqnarray} 
where in the second line we inverted~(\ref{cumu}) to express the PPT distribution as the derivative of the cumulative distribution. 
The integration of~(\ref{ppt222}) w.r.t.~$\tau$ {{ gives}} now (\ref{easyrecursive}) for $n=2$ concluding the proof.

Cases of special interests are those where the many-body system is formed by particles of the same type (say all qubits or qutrits) and {{the local noise is uniform}}. Under this condition Eqs.~(\ref{easyrecursive}) and (\ref{easyrecursive1}) leads to 
\begin{eqnarray} \label{easyrecursiveN} 
\bar{P}^{[{\rm loc},n]}_{ppt}(T) &=&  \left( \bar{P}^{(1)}_{ppt}(T)\right)^n\;,\\ 
 \label{easyrecursive1N} 
{P}^{[{\rm loc},n]}_{ppt}(\tau) &=&
n {P}^{(1)}_{ppt}(\tau) \left( \bar{P}^{(1)}_{ppt}(\tau)\right)^{n-1}  \;.
\end{eqnarray}
A direct consequence of~(\ref{easyrecursiveN}) is that 
for fixed value of $T$ the probability $\bar{P}^{[{\rm loc},n]}_{ppt}(T)$ is monotonically non-increasing w.r.t. to $n$, meaning that the 
entanglement in these  models tends to be degraded more slowly as the system size grows. 
This is to be expected, since it is more likely that there is always at least one of the systems involved which is entangled with its ancilla, so the time after which all systems become disentangled increases. 
A close inspection of the identities~(\ref{easyrecursiveN}) and (\ref{easyrecursive1N})  
 reveals finally that the ratio between the PPTT distribution and its cumulative counterpart is
extensive w.r.t.~to the number of sites, i.e. 
\begin{eqnarray} 
{R}^{[{\rm loc},n]}_{ppt}(\tau):= \frac{{P}^{[{\rm loc},n]}_{ppt}(\tau)}{\bar{P}^{[{\rm loc},n]}_{ppt}(\tau)} = n {R}^{(1)}_{ppt}(\tau)\;, 
\end{eqnarray} 
with ${R}^{(1)}_{ppt}(\tau):= {{P}^{(1)}_{ppt}(\tau)}/{\bar{P}^{(1)}_{ppt}(\tau)}$.

	{{\section{Dynamical Generator Sampling}\label{sec:samp}
	
In this section we study the PPTT distribution and its cumulative counterpart for some low dimensional systems
under the assumption of minimal amount of prior information on the noise model. Specifically, 
	maintaining}} the structure~(\ref{eq:Lindbladianalphagamma}) for the generator ${\cal L}$,
	we introduce dedicated ensembles for $\hat{H}$ and $K$ 
	that capture all the global characteristics of the system to be described, along with the Lindbladian invariances.
In the case of the Hamiltonian component
	  we sample $\hat{H}$ on the set of Hermitian matrix without symmetries constraints~\cite{NOTA2}: it can be proven that the natural ensemble for matrices with these characteristics, that are invariant under transformations of the $U(N)$ group, is the Gaussian Unitary Ensemble ($\mbox{GUE($N$)}$) \cite{Mehta}. For what concerns the dissipator component of (\ref{eq:Lindbladianalphagamma}) 
	having fixed the orthonormal basis $\{\hat{F}_{n}\}_{n=1,\dots,N^2-1}$ as the set of 
	infinitesimal generators of $SU(N)$~\cite{Zyczkowski}, 
we sample $D_K$ on an 
 ensemble of matrices $K$ which is invariant under unitary transformations (the Lindbladian is indeed invariant under changes of the basis of matrices and this property reflects in the invariance of the Kossakowski matrix). As explained in Ref.~\cite{LangeTimm}, to meet these requirements, one can consider the so-called Wishart ensemble, made up of matrices of the form $W=A^{\dagger}A\ge0$, where $A$ is a complex square matrix, sampled from the Ginibre Unitary Ensemble ($\mbox{GinUE($N$)}$), whose entries are complex with real and imaginary parts that are independently normally distributed. This choice ensures that, by construction, all the sampled matrices $K$ are positive semidefinite and that they are invariant under unitary transformations, as a consequence of the same invariance of the Ginibre ensemble~\cite{Mezzadri}.  Given hence $A\in\mbox{GinUE($N$)}$ we define its associated  Kossakowski matrix as
	$K(A):=N\frac{A^{\dagger}A}{\text{Tr}(A^{\dagger}A)}$,
	which by construction is dimensionless and has trace equal to $N$. It bears noting that, in this way, apart from the normalization condition, sampling $K(A)$ is equivalent to sample a random density matrix with a flat geometry~\cite{Bengtsson}.

Once the ensembles for sampling $\hat{H}$ and $K$ have been selected, we can examine the PPTT probability {{distributions $P^{(\alpha,\gamma)}_{ppt}(\tau)$~(\ref{mpodef})
which, for fixed values of $\alpha$ and $\gamma$, govern the statistics of the PPTTs}} associated with the generators $\mathcal{L}_{H,K}^{(\alpha,\gamma)}$ defined in Eq.~(\ref{eq:Lindbladianalphagamma}).
Reconstructing  $P^{(\alpha,\gamma)}_{ppt}(\tau)$ {{and its cumulative counterpart 
$\bar{P}^{(\alpha,\gamma)}_{ppt}(\tau)$~(\ref{cumu}),}}
 is generally a complex task as it involves numerically integrating the dynamics of the corresponding Choi-Jamiołkowski state~(\ref{choi}) for each sampled value of $\hat{H}$ and $K$, and determining the first time $t$ at which {{ it  becomes PPT}}. Consequently, in the following discussion, we will explicitly focus first on the case where $S$ is {{a qubit ($N=2$ where, regardless of the choice of the generator, 
$\tau_{ppt}({\cal L})$ and $\tau_{est}({\cal L})$ always coincide) 
or a qutrit $(N=3)$.}} 
A further simplification we shall adopt arises from the scaling (\ref{eq:scaling_law_beta}) which, setting 
$\beta=1/\gamma$,
implies
{{\begin{eqnarray}\label{IMPO}  
\begin{cases} 
P^{(\alpha,\gamma)}_{ppt}(\tau)  &= {\gamma} {\mathbf P}^{(k)}_{ppt}(\gamma
\tau) \;, \\\\ 
\bar{P}^{(\alpha,\gamma)}_{ppt}(\tau)  &=  \bar{\mathbf P}^{(k)}_{ppt}(\gamma
\tau) \;, 
\end{cases} 
\end{eqnarray} 
where $k:=\alpha/\gamma$ and
${\mathbf P}^{(k)}_{ppt}(x)$, $\bar{\mathbf P}^{(k)}_{ppt}(X):=\int_0^X {\mathbf P}^{(k)}_{ppt}(x)$  express 
the PPTT distribution and its cumulant  in $(1/\gamma)$-units corresponding to the (dimensionless)}}
	LGKS generator~\cite{NOTA2}
\begin{equation}{\label{eq:Lindbladian_riscalato}}
		\mathcal{L}_{H,K}^{(k,1)}(\,\cdots \,):= -i  k [\hat{H}, \cdots \,] +  \mathcal{D}_{K}(\,\cdots \,)\;.
	\end{equation} 
	Alternatively  by setting $\beta = 1/\alpha$ we can also write 
{{\begin{eqnarray}\label{IMPO1}  \begin{cases} 
P^{(\alpha,\gamma)}_{ppt}(\tau)  &= {\alpha} {\mathbf Q}^{(k)}_{ppt}(\alpha
\tau) \;,\\\\
\bar{P}^{(\alpha,\gamma)}_{ppt}(\tau)  &=  \bar{\mathbf Q}^{(k)}_{ppt}(\alpha
\tau)\;, 
\end{cases} 
\end{eqnarray} 
where now 
${\mathbf Q}^{(k)}_{ppt}(x)$ and $\bar{\mathbf Q}^{(k)}_{ppt}(X):=\int_0^X {\mathbf Q}^{(k)}_{ppt}(x)$  express 
the PPTT distribution and its cumulant  in $(1/\alpha)$-units  corresponding to the (dimensionless)}}
	LGKS generator~\cite{NOTA2}
\begin{equation}{\label{eq:Lindbladian_riscalato1}}
		\mathcal{L}_{H,K}^{(1,1/k)}(\,\cdots \,):= -i  [\hat{H}, \cdots \,] + (1/k)  \mathcal{D}_{K}(\,\cdots \,)\;.
	\end{equation} 
	It goes without mentioning that the functions {{(\ref{IMPO}) and (\ref{IMPO1})  are linked by the identity
	\begin{eqnarray}\begin{cases} 
	{\mathbf P}^{(k)}_{ppt}(x) = {k} {\mathbf Q}^{(k)}_{ppt}(kx)\;, \\\\
	\bar{\mathbf P}^{(k)}_{ppt}(X) =  \bar{\mathbf Q}^{(k)}_{ppt}(kX)\;.
	  \end{cases} 
	\end{eqnarray}}} 
	{{\section{Empirical distributions}\label{sect:ESTs_distrib}}}
	
	\begin{figure*}
		\subfloat[][]
		{\includegraphics[scale=0.225]{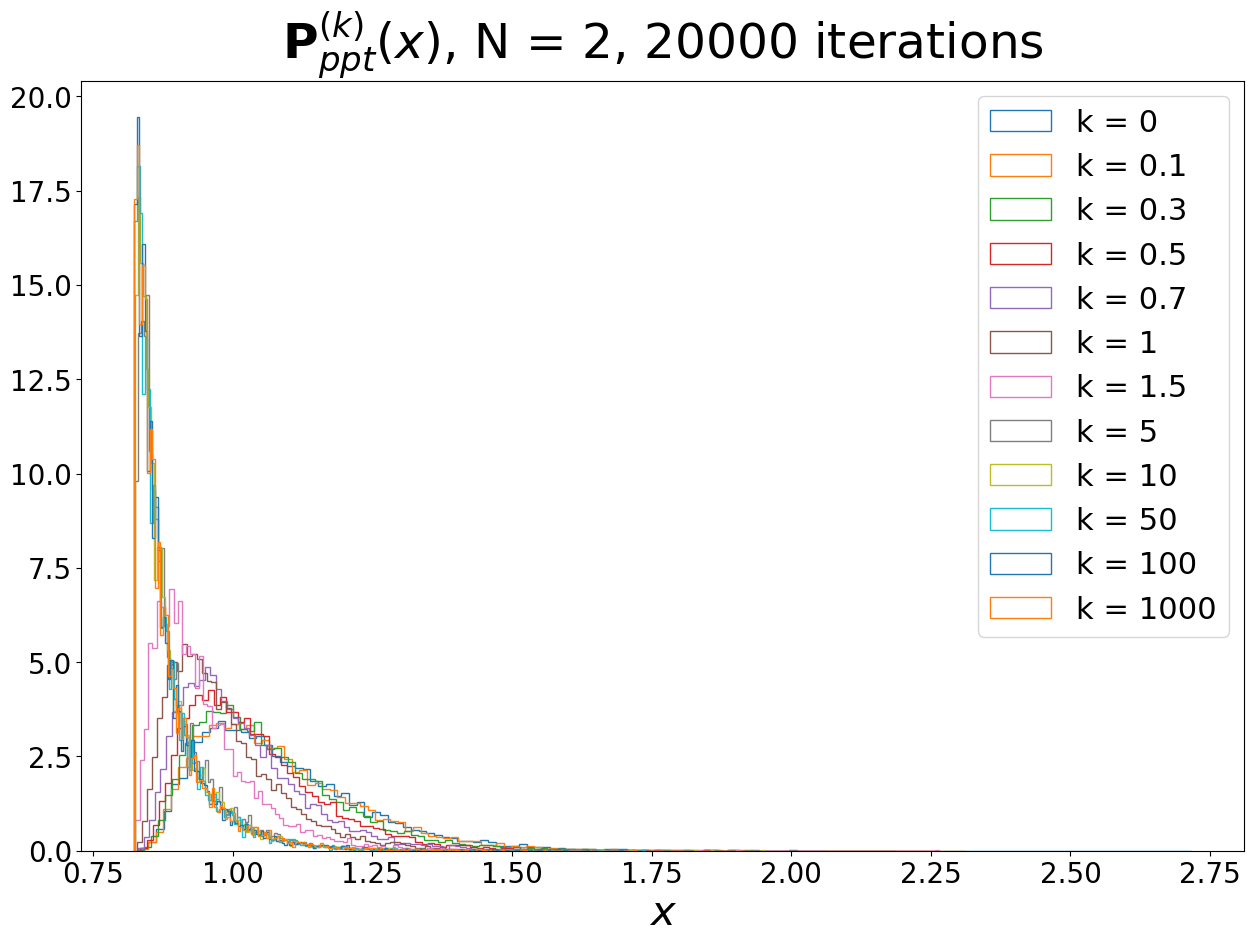}\label{fig:hist_panel_a}}\quad \quad
		\subfloat[][]
		{\includegraphics[scale=0.225]{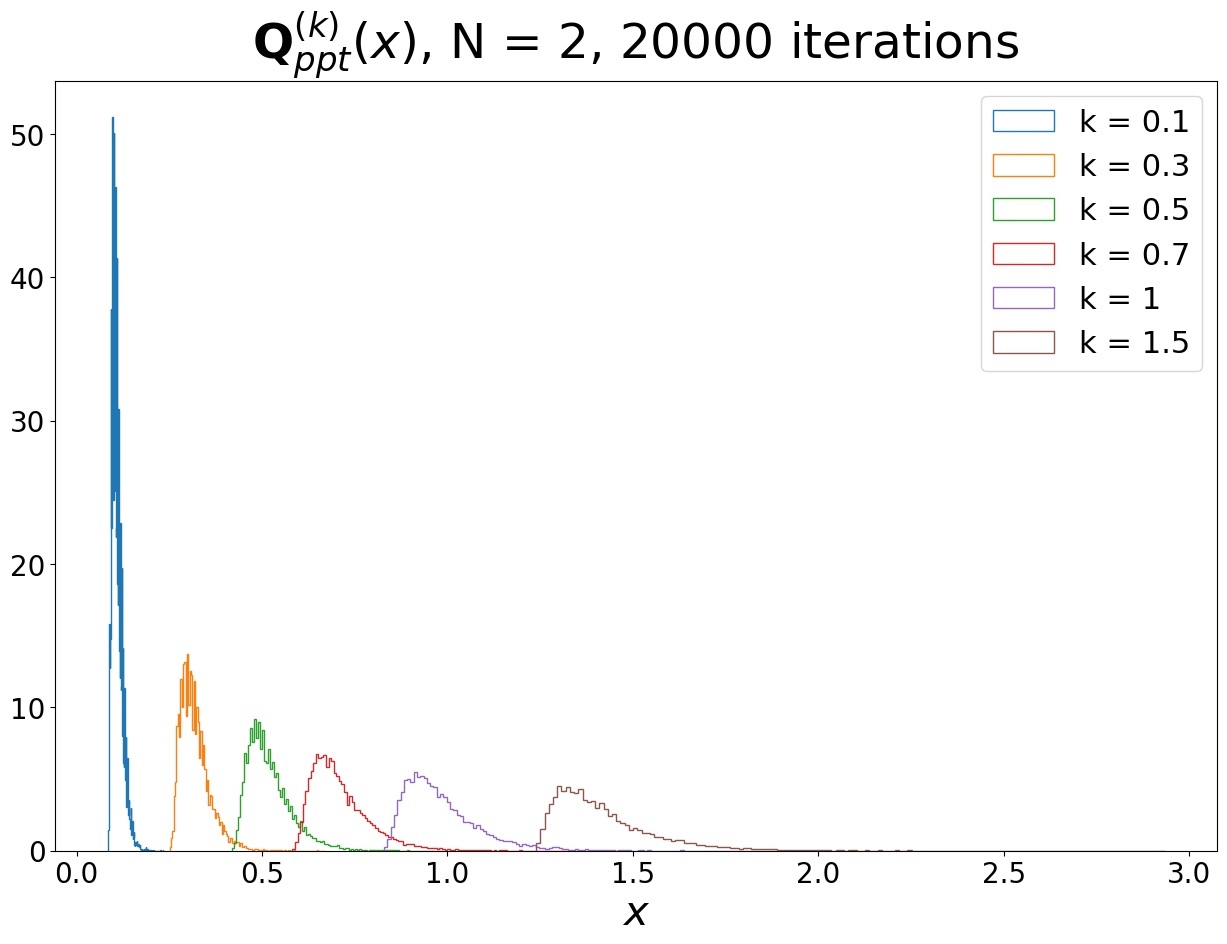}\label{fig:hist_panel_b}}
		\caption{{{Histograms of the empirical PPTT distributions 
		 for  a qubit ($N = 2$) system  for different instances
			of the rescaled Lindbladians $\mathcal{L}_{H,K}^{(k,1)}$ and $\mathcal{L}_{H,K}^{(1,1/k)}$. For each assigned value of $k$ we used 20000 independent   sampling of  $\hat{H}$ in $\mbox{GUE($N=2$)}$ and $A$ with  $\mbox{GinUE($N=2$)}$. Panel (a): histograms associated to the rescaled distribution  ${\mathbf P}^{(k)}_{ppt}(x)$ of Eq.~(\ref{IMPO}) where  $\gamma$ is fixed and different values of $\alpha= k\gamma$ are considered. Notice the emergence of a non trivial limit distribution ${\mathbf P}^{(\infty)}_{ppt}(x)$ as $k$ increases.		Panel (b): histograms associated to the rescaled distribution ${\mathbf Q}^{(k)}_{ppt}(x)$ of Eq.~(\ref{IMPO1}) 	where  instead $\alpha$ is fixed and different values of $\gamma = \alpha/k$ are considered.}}}
		\label{fig:Histrograms}
		\bigbreak
		\subfloat[][]
		{\includegraphics[scale=0.225]{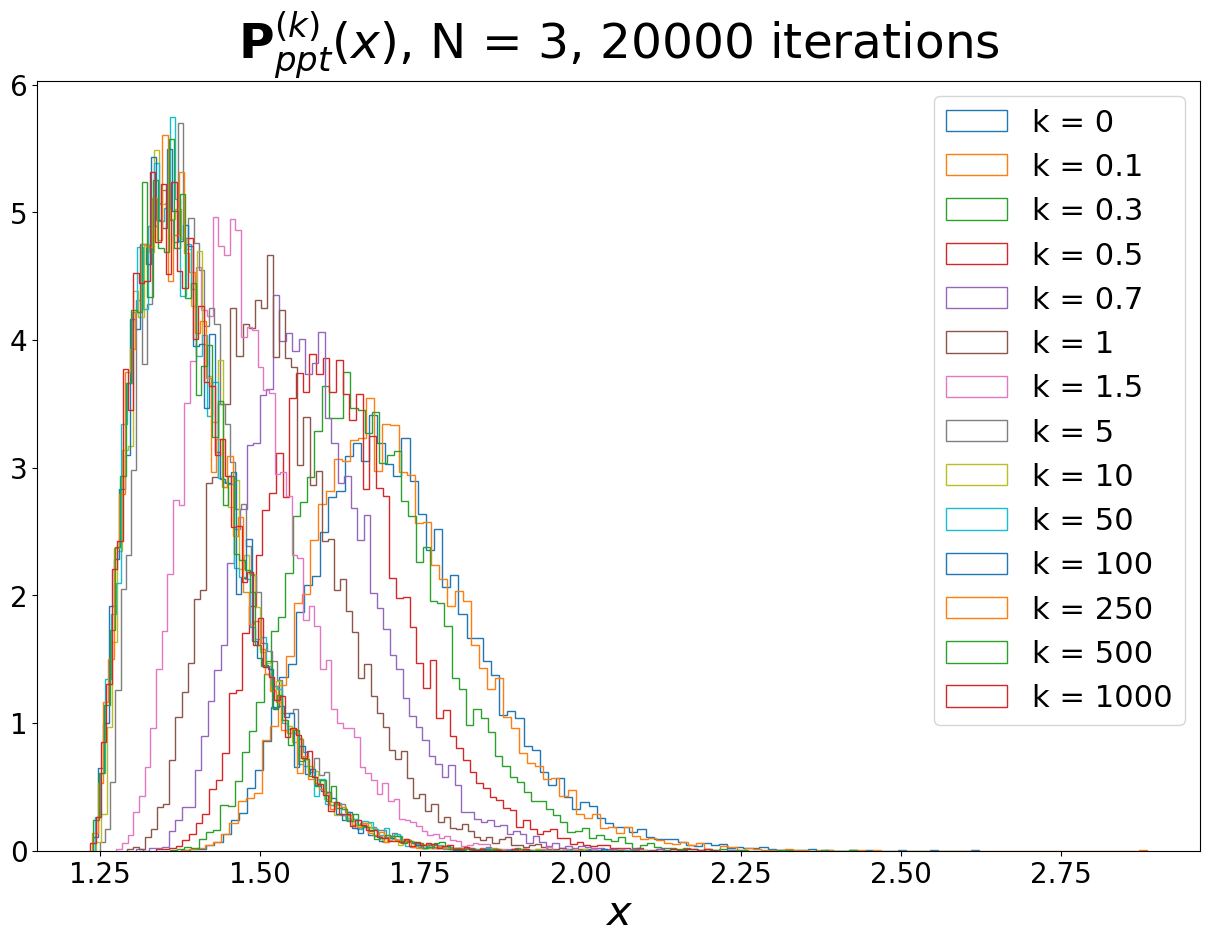}\label{fig:histN3_panel_a}}\quad \quad
		\subfloat[][]
		{\includegraphics[scale=0.225]{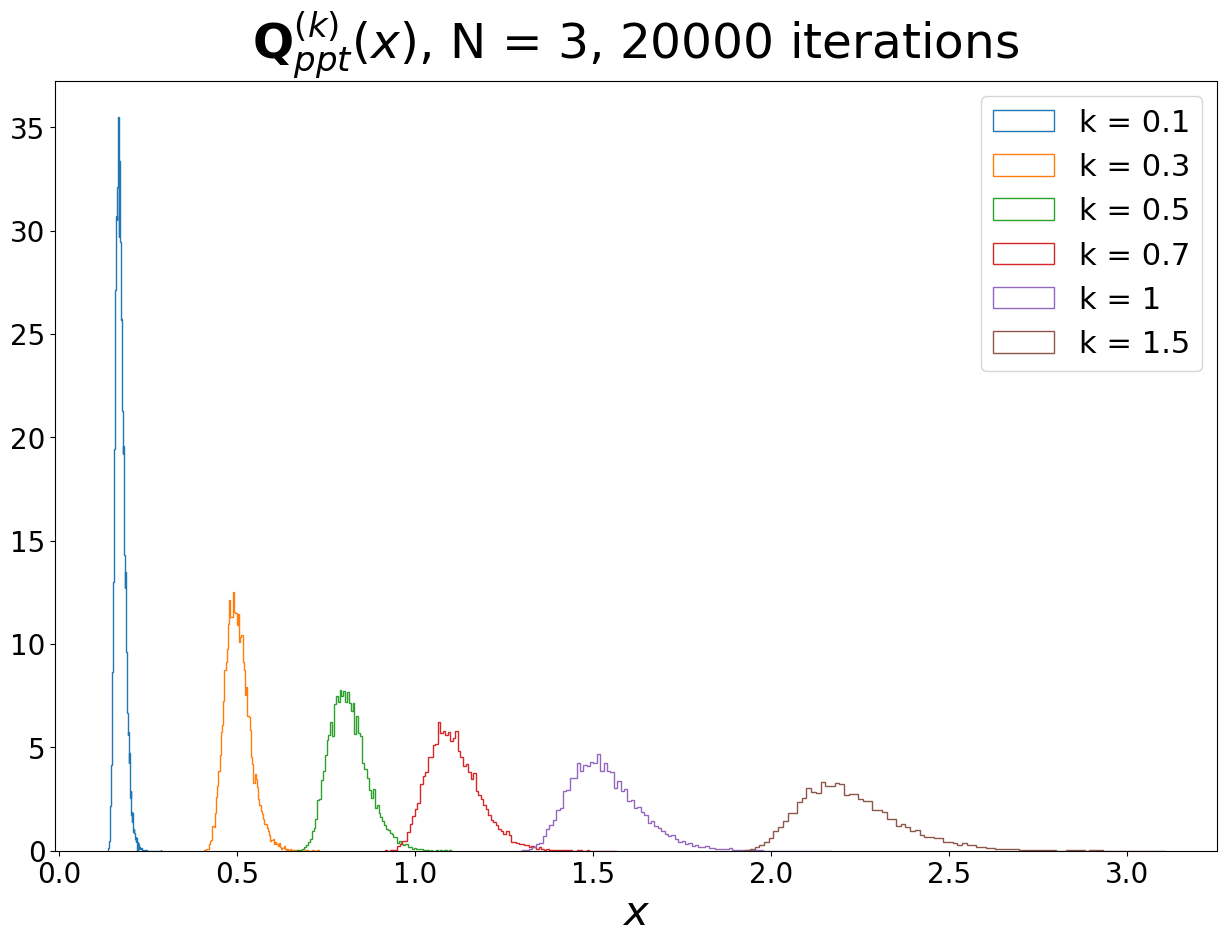}\label{fig:histN3_panel_b}}
		\caption{{{Same plots as Fig.~\ref{fig:Histrograms} for a qutrit ($N=3$) system. Also in this case we
		used 20000 independent sampling of $\hat{H}$ in  $\mbox{GUE($N=3$)}$ and $A$ with  $\mbox{GinUE($N=3$)}$. }}
			}
		\label{fig:Histrograms_N3}
	\end{figure*}
	
	\begin{figure*}
		\subfloat[][]
		{\includegraphics[scale=0.31]{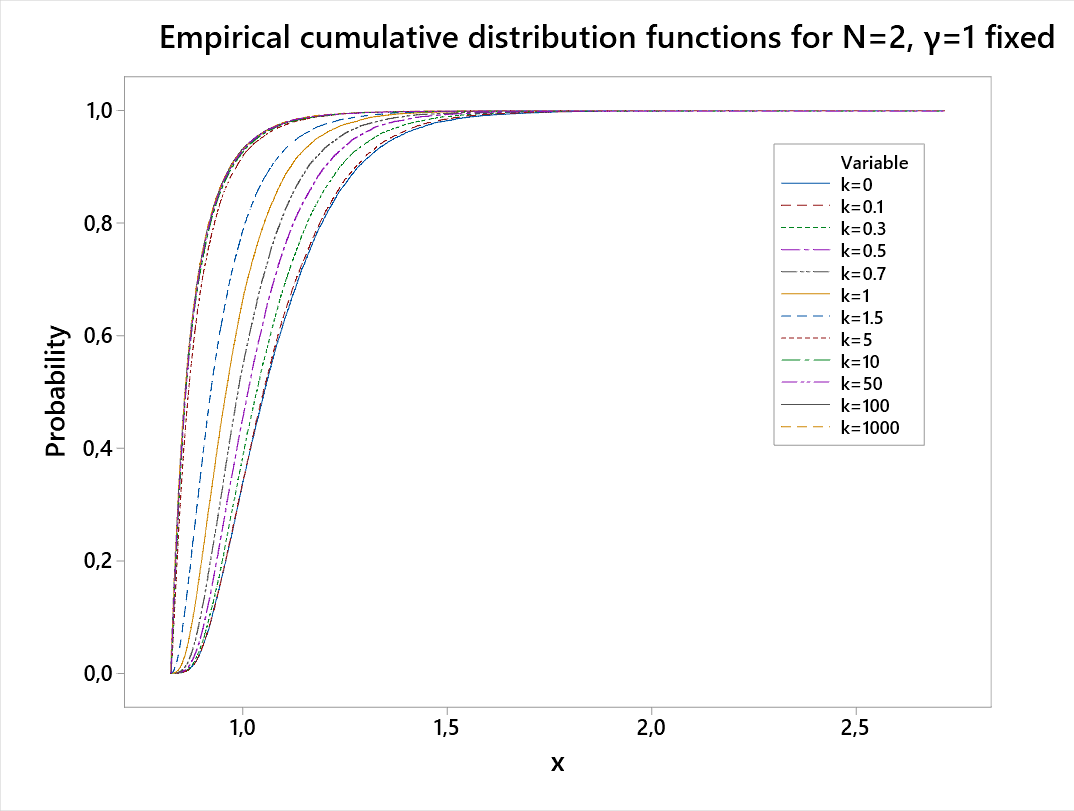}\label{fig:cdf_N=2_panel_a}}\quad \quad
		\subfloat[][]
		{\includegraphics[scale=0.31]{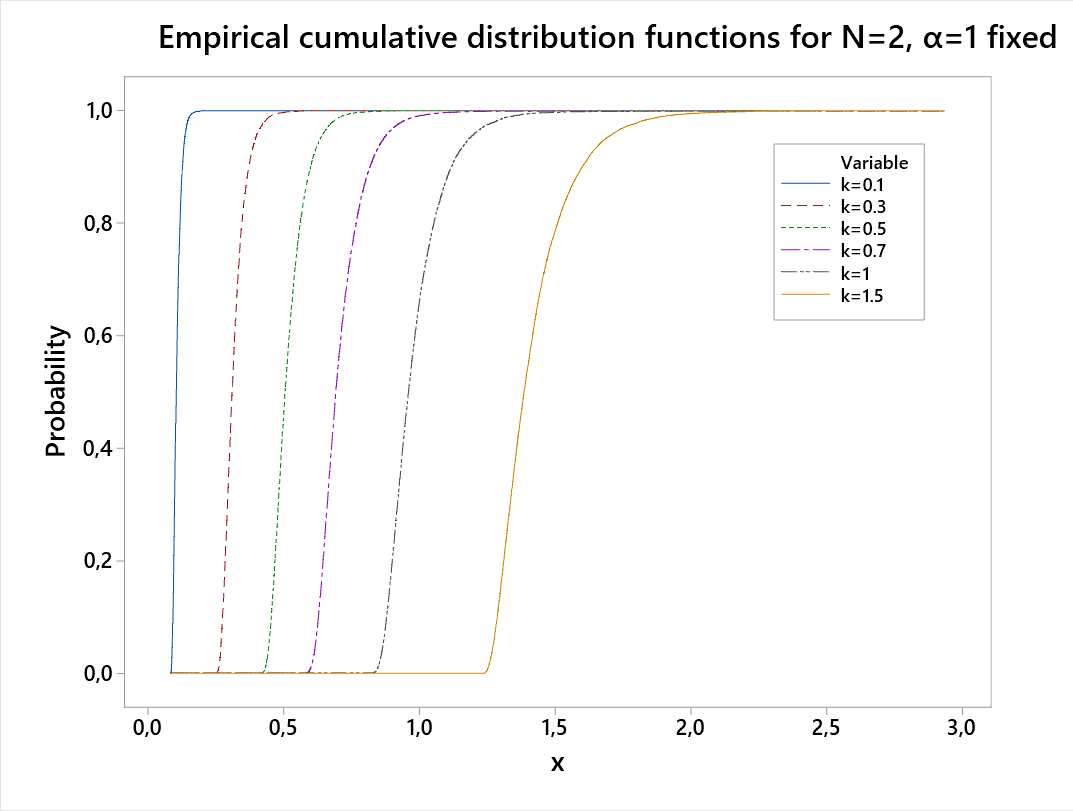}\label{fig:cdf_N=2_panel_b}}
		\caption{{{Empirical cumulative distribution functions related to the PPTT distributions of Figs.~\ref{fig:hist_panel_a} and \ref{fig:hist_panel_b} for a qubit ($N = 2$) system for different instances of the rescaled Lindbladians $\mathcal{L}_{H,K}^{(k,1)}$ and $\mathcal{L}_{H,K}^{(1,1/k)}$. Panel (a): empirical cumulative distribution function $\bar{{\mathbf P}}^{(k)}_{ppt}(x)$ of Eq.~(\ref{IMPO}) where  $\gamma$ is fixed and different values of $\alpha= k\gamma$ are considered.	Panel (b): empirical cumulative distribution function $\bar{{\mathbf Q}}^{(k)}_{ppt}(x)$ of Eq.~(\ref{IMPO1}) where instead $\alpha$ is fixed and different values of $\gamma = \alpha/k$ are considered.}}}
		\label{fig:cdf_N=2}
		\bigbreak
		\subfloat[][]
		{\includegraphics[scale=0.31]{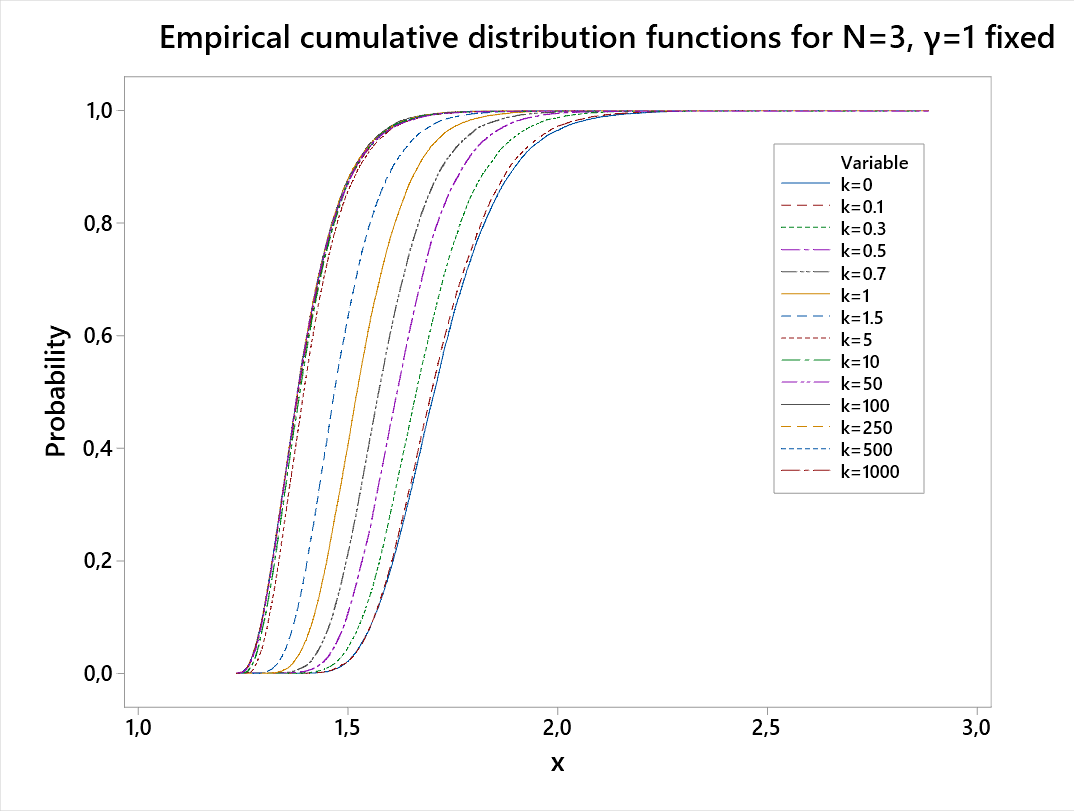}\label{fig:cdf_N=3_panel_a}}\quad \quad
		\subfloat[][]
		{\includegraphics[scale=0.31]{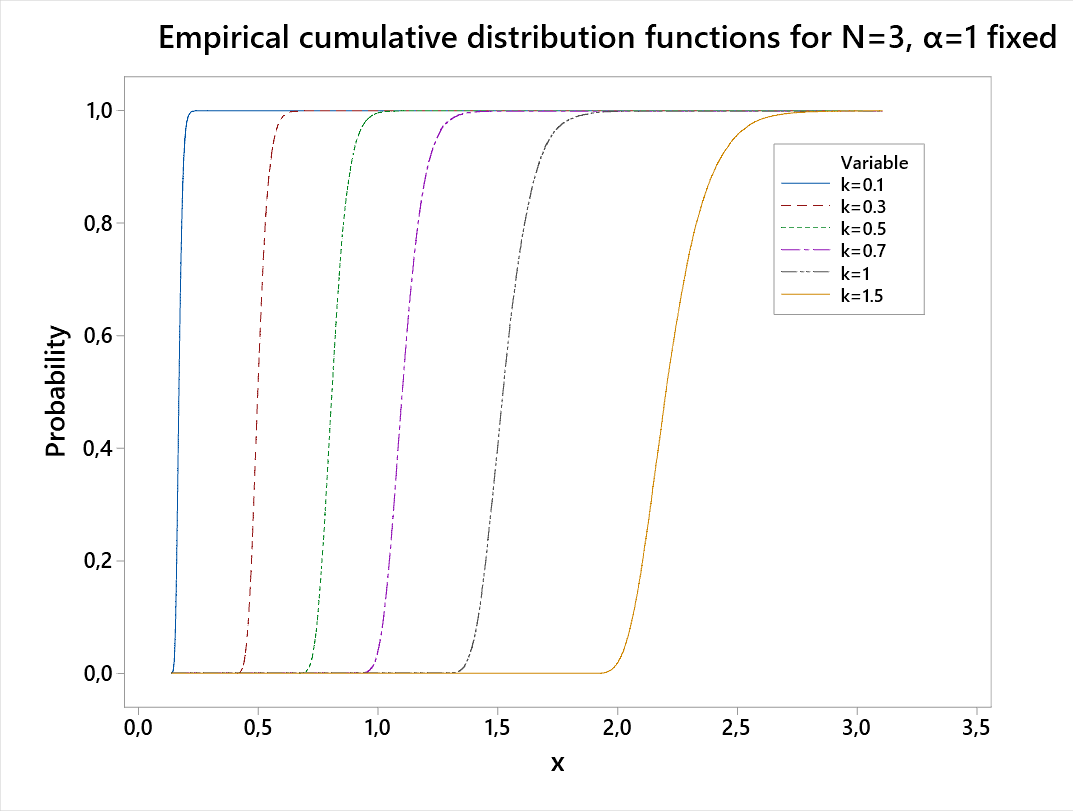}\label{fig:cdf_N=3_panel_b}}
		\caption{{{Same plots as Fig.~\ref{fig:cdf_N=2} for a qutrit ($N=3$) system.}}
		}
		\label{fig:cdf_N=3}
	\end{figure*}

	{{In this section we report  the empirical estimation of the rescaled PPTT distributions 
	 ${\mathbf P}^{(k)}_{ppt}(x)$ and ${\mathbf Q}^{(k)}_{ppt}(x)$ and their cumulative counterpart $\bar{\mathbf P}^{(k)}_{ppt}(x)$ and $\bar{\mathbf Q}^{(k)}_{ppt}(x)$,  for different values of the coefficient $k$ in the case of qubit and qutrit quantum system $(N=2,3)$, obtained by 
	 sampling the corresponding Lindblad generators  $\mathcal{L}_{H,K}^{(k,1)}$ and $\mathcal{L}_{H,K}^{(1,1/k)}$ 
	 through the uniform measure defined in the previous section. 
	For each simulation, we sampled $20000$  Lindbladians 
	 and derived {{the PPTTs}} associated with them. The number of samplings chosen is justified by the fact that it was sufficient to reveal the shape of the distributions while still guaranteeing acceptable computational performances in terms of time.

	 Figures ~\ref{fig:Histrograms}  and~\ref{fig:Histrograms_N3} showcase the empirical PPTT distributions derived for $N = 2$ and $N = 3$, respectively.
In particular the panels (a) of these figures report the histograms  associated with
${\mathbf P}^{(k)}_{ppt}(x)$ of Eq.~(\ref{IMPO}) where all the quantities are expressed in units of~(1/$\gamma$). The plots reveal that ${\mathbf P}^{(k)}_{ppt}(x)$
 become more and more peaked as $k$ increases  and a  non trivial limit distribution ${\mathbf P}^{(\infty)}_{ppt}(x)$ appears to emerge for   $k\rightarrow\infty$ (see also  Sec.~\ref{sect:LimLind} below).  
These facts  can be interpreted as a generalization of what was observed in~\cite{Gatto} and it indicates that, at least at the distribution level, for a dissipative contribution of fixed strength (in this case $\gamma=1$), increasingly strong unitary terms in the Lindbladian behave as an additional noise that tends to cause entanglement to vanish sooner. 
	The plots of the panels~(a) seem also to suggest that all the histograms  have the same minimum time. However, from a subsequent analysis, it emerges that there exists a relation that links such times to the Lindbladian parameter $k$, which reveals that they have a small range of variation. Observe finally that  when comparing the qubit case with the qutrit case, it is apparent that these latter distributions are characterized by longer entanglement degradation times as they are shifted to the right if compared to the former ones. 
Panels~(b) of  Figs.~\ref{fig:Histrograms}  and~\ref{fig:Histrograms_N3} represent instead the histograms 
 obtained  by samplings the rescaled Lindbladian $\mathcal{L}_{H,K}^{(1,1/k)}$ 
of (\ref{eq:Lindbladian_riscalato1}) for different values of $k$ --  the associated
 theoretical 
	 distribution being the function {{${\mathbf Q}^{(k)}_{ppt}(x)$}} of~(\ref{IMPO1}) where everything is expressed in units of (1/$\alpha$). 
	What emerges in this case is in line with the expectations: in fact, once having fixed the strength of the unitary Hamiltonian term (in this case $\alpha=1$ according to Eq.~(\ref{eq:Lindbladian_riscalato1})), the limiting case $k\ll 1$ represents the physical condition of a very noisy environment ($\gamma \gg 1$)  which causes the system to disentangle rapidly; on the contrary  for {{ $k\gg 1$
	(i.e. $\gamma \ll 1$)}} entanglement vanishes at increasingly longer times as the system becomes more and more similar to a closed system.
		
	In Figs.~\ref{fig:cdf_N=2} and \ref{fig:cdf_N=3} we finally report the plots of the empirical cumulative distribution functions. Specifically, in the panels (a) of these figures  
	we present the functions $\bar{{\mathbf P}}^{(k)}_{ppt}(x)$ of Eq.~(\ref{IMPO}), associated respectively with the PPTT distributions of the qubit ($N=2$) and qutrit ($N=3$) case respectively, 
	(i.e.  the cumulative integrals of the histograms of 
	reported in Figs.~\ref{fig:Histrograms}(a)  and~\ref{fig:Histrograms_N3}(a)
	 and obtained by sampling the generator ${\cal{L}}_{H,K}^{(k,1)}$ for different values of $k$). Analogously, in the panels (b)  we present the functions $\bar{{\mathbf Q}}^{(k)}_{ppt}(x)$ of Eq.~(\ref{IMPO1}) associated with the PPTT distributions reported in panel
	 Figs.~\ref{fig:Histrograms}(b)  and~\ref{fig:Histrograms_N3}(b)
and related to the generator  ${\cal{L}}_{H,K}^{(1,1/k)}$.
}}

	\subsection{Limit Lindbladian}\label{sect:LimLind}
	From the plots reported in Figs.~\ref{fig:hist_panel_a},~\ref{fig:histN3_panel_a} 
	it emerges that when the system exhibits dynamics characterized by a unitary Hamiltonian contribution that is predominant with respect to the dissipative term, the {{rescaled distributions ${\mathbf P}^{(k)}_{ppt}(x)$ tend to a limit distribution ${\mathbf P}^{(\infty)}_{ppt}(x)$,}} through a process which is distinguished by mean and median times that got smaller and smaller as $k$ increases. We are now interested in finding the explicit expression of the 
	effective Lindbladian $\mathcal{L}_{H,K}^{({\rm{eff}})}$ that, properly sampled over $\hat{H}$ and $K$, allows one to obtain the limit distribution {{${\mathbf P}^{(\infty)}_{ppt}(x)$}}. We will refer to this super-operator as \textit{limit Lindbladian}. 
	
	As a starting point of our analysis let us express the GKSL master equation associated with  {{the generator $\mathcal{L}_{H,K}^{(k,1)}$}} of Eq.~(\ref{eq:Lindbladian_riscalato}) 
	in interaction picture w.r.t. to the Hamiltonian component, i.e. 
	\begin{equation}
		\dot{\tilde{\rho}}_{k}(t) = \tilde{\cal L}_{H,K}^{(kt)} (\tilde{\rho}_{k}(t)),
	\end{equation}
	where 
	\begin{eqnarray}{\label{eq:operators_inter_pict}}
				 \tilde{\rho}_{k}(t) &:=& e^{ik \hat{H} t} \hat{\rho}(t) e^{-i k \hat{H} t}, 
				 \end{eqnarray}
and $\tilde{\cal L}_{H,K}^{(kt)}$ the super-operator obtained from 
 Eq.~(\ref{eq:Dissipator_Kmn}) 			 by replacing 
 the operators $\hat{F}_n$ with 
 \begin{eqnarray} 
\tilde{F}_{n; H}^{(kt)} &:=& e^{ik \hat{H} t} \hat{F}_n e^{-i k \hat{H} t}\;.
 \end{eqnarray} 
{{Notice that in the above expressions, for the sake of simplicity we dropped the index $S$. Notice also that since
$\tilde{\rho}_{k}(t)$ and $\hat{\rho}(t)$  are connected via a unitary transformations that act locally on the system of interests, their corresponding Choi-Jamiołkowski  states  share the same entanglement properties, meaning that PPTT values associated with
$\tilde{\cal L}_{H,K}^{(kt)}$ exactly matches those of the original Lindbladian $\mathcal{L}_{H,K}^{(k,1)}$.}}
	An ansatz for the limit Lindbladian is obtained by considering the asymptotic average of   ${\cal D}_{K;H}^{(kt)}$ for large $k$, i.e. 
	\begin{eqnarray}
		\mathcal{L}_{H,K}^{({\rm{eff}})} &:=&  \lim_{k \rightarrow \infty} \frac{1}{k} \int_{0}^{k} d k' \; \tilde{\cal L}_{H,K}^{(k't)}\;,
			\label{eq:limit_lindbladian_def}
	\end{eqnarray}	
	(as we shall see, as long as $t>0$, the resulting super-operator does not bare any functional dependence upon such parameter). The resulting superoperator no longer depends on $k$, but only on the  eigenvectors of the selected Hamiltonian, as well as the original Kossakowski matrix $K$. To perform the integral in Eq.~\eqref{eq:limit_lindbladian_def}, we divide it into three terms and compute them separately. Under these assumptions, the second and the third term of the integral, given by the anticommutator in the Lindblad master equation, will give the same result and one can compute only one of the two.
	Leaving the explicit derivation in Appendix~\ref{app:Limit_Lindbladian}, we report here the  final expression, written in terms of the Lindblad operators  $\{{\hat{L}^{(n)}_K}\}_{n=1,\cdots, N^2-1}$ of the dissipator ${\cal D}_K$: 
		\begin{widetext}
		\begin{equation}
			\mathcal{L}_{H,K}^{({\rm{eff}})}= \sum_{n = 1}^{N^2 - 1}  \Bigg[\sum_{i,j}\hat{Z}_{i,i}^{(n)}(\cdots) \hat{Z}_{j,j}^{(n)\, \dagger} + \sum_{i,j\neq i} \hat{Z}_{i,j}^{(n)}(\cdots) \hat{Z}_{i,j}^{(n)\, \dagger} - \frac{1}{2} \sum_{i, j} \Big( {\hat{Z}_{j,i}^{(n)\,\dagger}}  \hat{Z}_{j,i}^{(n)}(\cdots) + (\cdots) {\hat{Z}_{j,i}^{(n)\,\dagger}} \hat{Z}_{j,i}^{(n)} \Big)\Bigg] \;,
			\label{eq:L_lim_Z}
		\end{equation}
	\end{widetext}
	where indicating with $\hat{\Pi}_i$ the projector on the $i$-th eigenspace of   $\hat{H}$ we define
	\begin{eqnarray}
		\hat{Z}_{i,j}^{(n)} 
			&:=&{\hat{\Pi}_i} {\hat{L}^{(n)}_K}{\hat{\Pi}_j} \;. 
	\end{eqnarray}
To cast Eq.~(\ref{eq:L_lim_Z}) in a more familiar form 
we perform an index conversion in $\hat{Z}_{i,j}^{(n)}$ operators, in order to identify them with one index. More specifically, one can do the following conversion $(i,j) \rightarrow m$, with $m=iN + j$.
	In this way, Eq.~\eqref{eq:L_lim_Z} can be written in terms of a sparse symmetric matrix $A$, whose elements are $0$ and $1$, and are determined in such a way to reproduce that equation, i.e.
	  \begin{figure*}[!htb]
	\subfloat[][]
	{\includegraphics[scale=0.22]{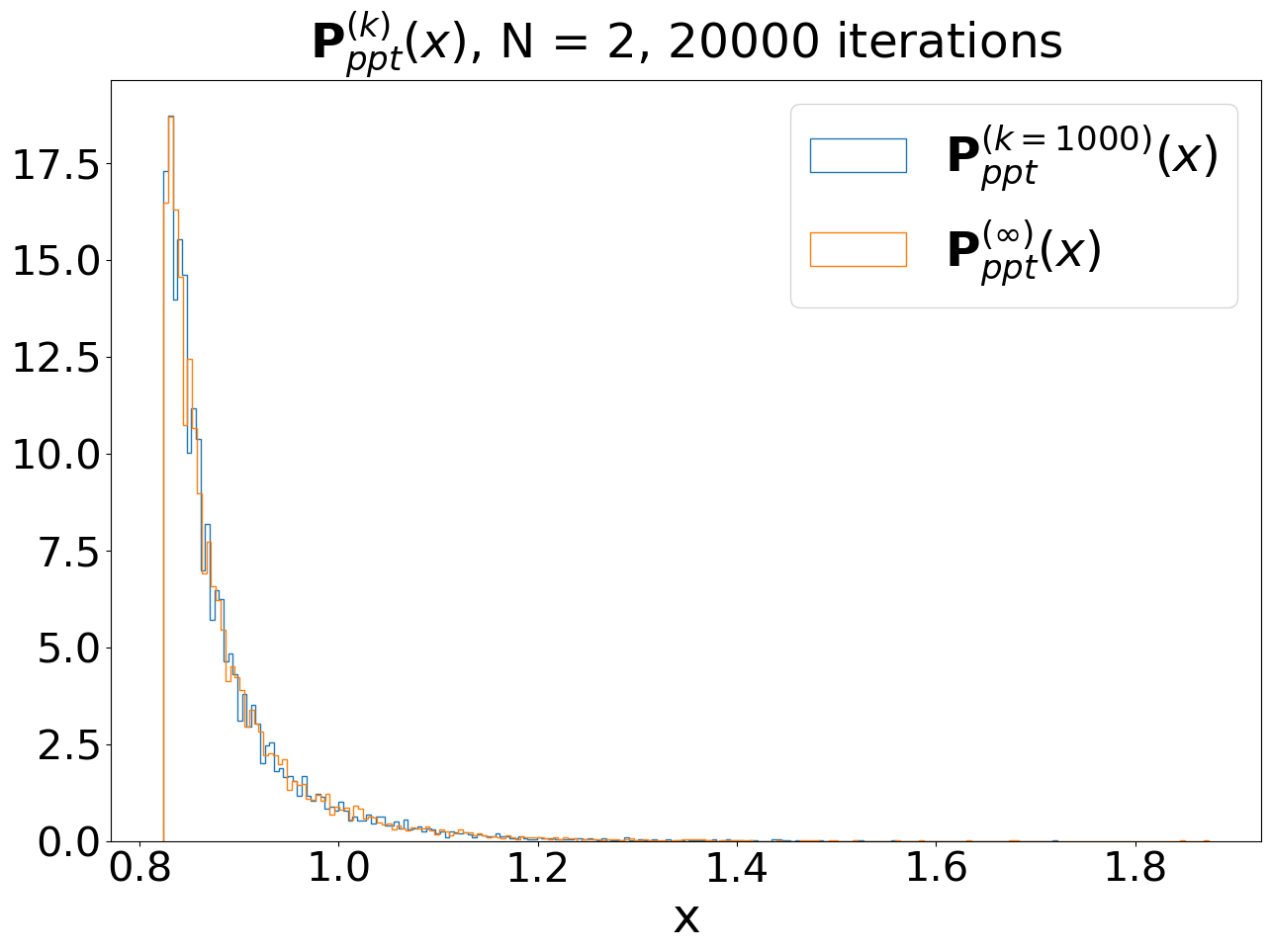}\label{fig:limit_overlap_N2}}\qquad
	\subfloat[][]
	{\includegraphics[scale=0.22]{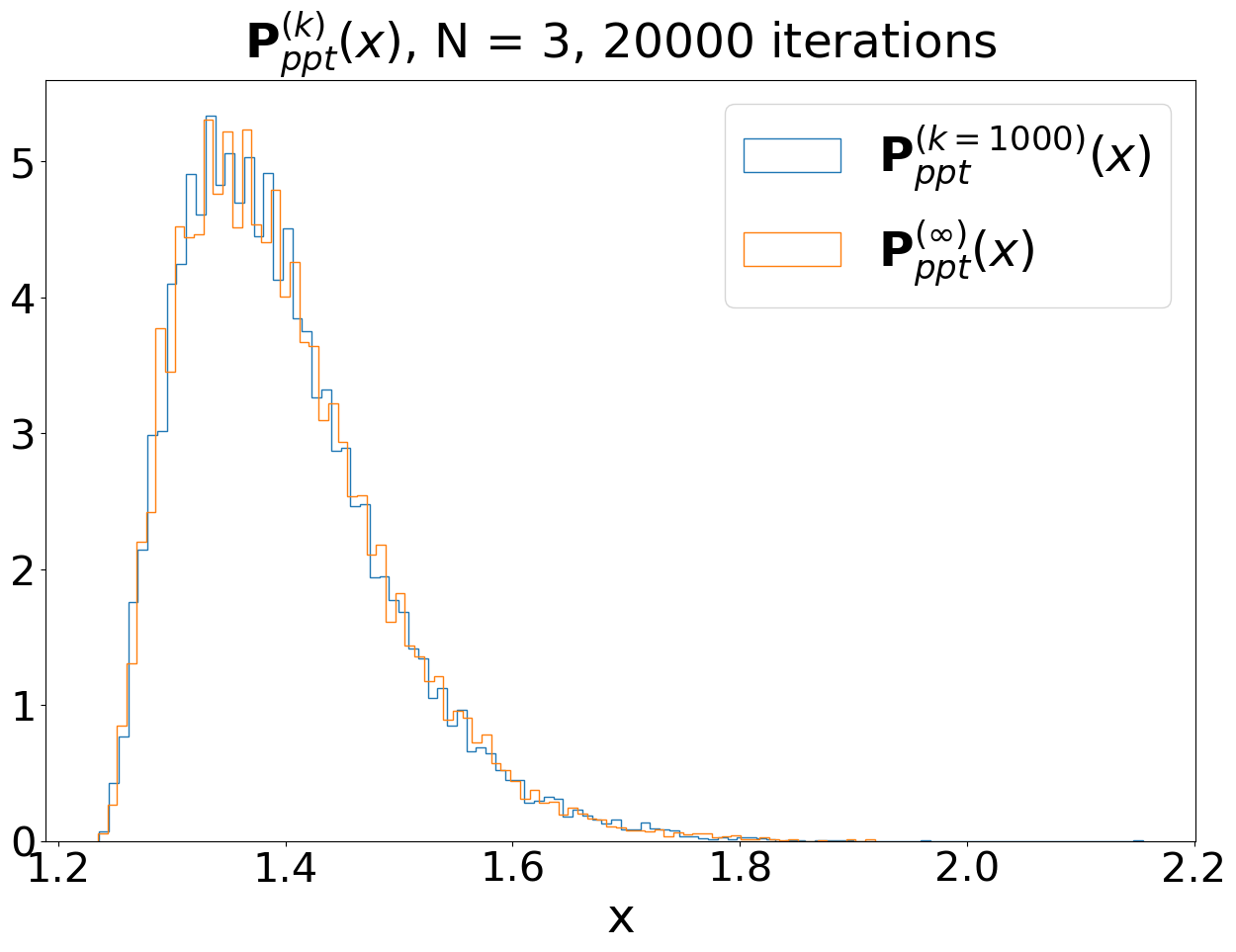}\label{fig:limit_overlap_N3}}
	\caption{Comparison between the asymptotic distribution simulated sampling the limit Lindbladian $\mathcal{L}_{H,K}^{({\rm{eff}})}$ and the distributions obtained by sampling the rescaled Lindbladian $\mathcal{L}_{H,K}^{(k,1)}$  evaluated for  $k=1000$. {{The comparison between ${\mathbf P}^{(\infty)}_{ppt}(x)$ and ${\mathbf P}^{(k=1000)}_{ppt}(x)$ for $N=2$ is reported in Panel (a) while that for $N=3$ in Panel (b).}}}
	\label{fig:lim_distrib_overlapped}
\end{figure*}
		\begin{eqnarray}\nonumber 
				&&\!\!\!\mathcal{L}_{H,K}^{({\rm{eff}})}=  \sum_{n = 1}^{N^2 - 1} \sum_{m,m'=0}^{N^2 - 1}  \, A_{mm'} \Bigg[\hat{Z}_{m}^{(n)}(\cdots) \hat{Z}_{m'}^{(n)\, \dagger}\\ &&\qquad  - \frac{1}{2} \Big( {\hat{Z}_{m'}^{(n){\dagger} }} \hat{Z}_{m}^{(n)}(\cdots)  + (\cdots) {\hat{Z}_{m'}^{(n){\dagger} }}\hat{Z}_{m}^{(n)} \Big) \Bigg].\nonumber \\
			\label{eq:L_lim_with_A}
		\end{eqnarray}
	In particular $A_{mm'}=1$ if $m'=m $ and $m'\neq m$ with $m,m'$ s.t. $(i_{m}=l_{m}) \land (i_{m'}=l_{m'})$, where $i_{m}\, (l_{m})$ is the first (second) index before the conversion, associated with $m$. From this expression it is clear that the matrix $A$ is symmetric, therefore diagonalizable. Defining as $V$ the unitary matrix whose columns are the eigenvectors of $A$ and as $\eta_s$ its eigenvalues, it is possible to introduce new Lindblad operators via the identity 
	\begin{equation}
			\hat{Y}^{(s,n)} = \sum_{m=0}^{N^2 -1} \sqrt{\eta^{(s)}}V^{(m,s)}\hat{Z}_{m}^{(n)},
			\label{eq:Lim_lind_Lindbald_matrices}
	\end{equation}
	and write Eq.~\eqref{eq:L_lim_Z} in Lindblad form, in terms of them, i.e.
	\begin{eqnarray}\label{eq:L_lim_lindblad_form}
		&&\!\!\!\mathcal{L}_{H,K}^{({\rm{eff}})} = \sum_{\ell = 1}^{N^2(N^2 - 1)}  \Bigg[\hat{Y}^{(\ell)}(\cdots)\hat{Y}^{(\ell)\dagger} \\ 
		&&\qquad  - \frac{1}{2} \Big( \hat{Y}^{(\ell)\dagger}\hat{Y}^{(\ell)}(\cdots)  + (\cdots) \hat{Y}^{(\ell)\dagger}\hat{Y}^{(\ell)}\Big) \Bigg]\;, \nonumber 
	\end{eqnarray}
	where $\ell$ now stands for the joint indexes $s$ and $n$ appearing in the expression of the new Lindblad operators \eqref{eq:Lim_lind_Lindbald_matrices}. 
	It is worth stressing that the $\hat{Y}^{(s,n)}$'s inherit from the
 $\hat{Z}_{m}^{(n)}$'s an implicit functional dependence upon the Kossakowski matrix $K$ and of the Hamiltonian $\hat{H}$ of the original Lindbladian ${\cal L}$.

 To verify the correctness of the derivation  we have constructed a new algorithm that, starting from the expression of the limit Lindbladian defined in Eq.~\eqref{eq:L_lim_Z}, and sampling properly both the Hamiltonian and the dissipator, generates the associated distribution of the entanglement survival times and positive partial transpose times.  In the Fig.~\ref{fig:lim_distrib_overlapped} we report the plot obtained when considering both the limit distribution and the ones in correspondence with the largest $k$ value considered in Figs.~\ref{fig:hist_panel_a},~\ref{fig:histN3_panel_a}: as clear from the figures these curves 
perfectly overlap (actually we have checked that there is good agreement already starting from $k=5$). This is in accordance with what is expected and confirms that the formal expression of the limit Lindbladian, reported in Eq.~\eqref{eq:L_lim_Z}, gives a good estimate of the limit distribution, obtained when the unitary Hamiltonian contribution becomes predominant.

{{
	\subsection{Entanglement degradation of a collection of $n$ qubits under Random Local Noise} \label{sec:ppt-local-results}
	
	In Sec.~\ref{sec:ppt-local} we have seen 
	that the cumulative probability function  for quantum memories 
	of many isodimensional quantum systems on which uniform and local random Markov noise acts (see Fig.~\ref{figmany1})  is obtained by raising the cumulative distribution of a single system to the number of systems involved. 
	Suppose hence to have a quantum memory of this kind, made up of $n$ qubit systems $(N=2)$ and suppose that each qubit is subjected to a random Markovian noise identified by a generator ${\cal{L}}_{H,K}^{(k,1)}$ with fixed value of $k$. 
According to
	Eq.~(\ref{easyrecursiveN}) the cumulative probability distribution of the model can be expressed as
	\begin{eqnarray}
	\bar{\mathbf P}^{[{\rm loc},n](k)}_{ppt}(X) = \left(\bar{\mathbf P}^{(k)}_{ppt}(X) \right)^n\;, 
	\end{eqnarray}  with 
	$\bar{\mathbf P}^{(k)}_{ppt}(X)$ the single site cumulative probability that we have empirically estimated in Fig.~\ref{fig:cdf_N=2}. 
	Fixing $k=1000$ (i.e. equal to biggest value considered in our statistical analysis) 
	  we report the resulting function in  Fig.~\ref{fig:cdf_n} for different choices of $n$. 
As expected, for fixed $X$, the function $\bar{\mathbf P}^{[{\rm loc},n](k)}_{ppt}(X)$ is monotonically 
decreasing w.r.t. $n$. To get a quantitative estimation of this effect we determine numerically  the 
time $X_n$ at which the function $\bar{\mathbf P}^{[{\rm loc},n](k)}_{ppt}(X)$ assumes value $1/2$, i.e. the quantity 
\begin{equation} X_{n} := 	\left(\bar{\mathbf P}^{[{\rm loc},n](k)}_{ppt}\right)^{-1}(1/2)=\label{lastlast} 
{{\bar{\mathbf P}}_{ppt}^{(k)^{-1}} } (\sqrt[n]{1/2})\;, \end{equation}  
which formally corresponds to the evaluation of the median of the PPTT distribution  
${\mathbf P}^{[{\rm loc},n](k)}_{ppt}(x)$.
	\begin{figure}[t]
		\includegraphics[scale=0.225]{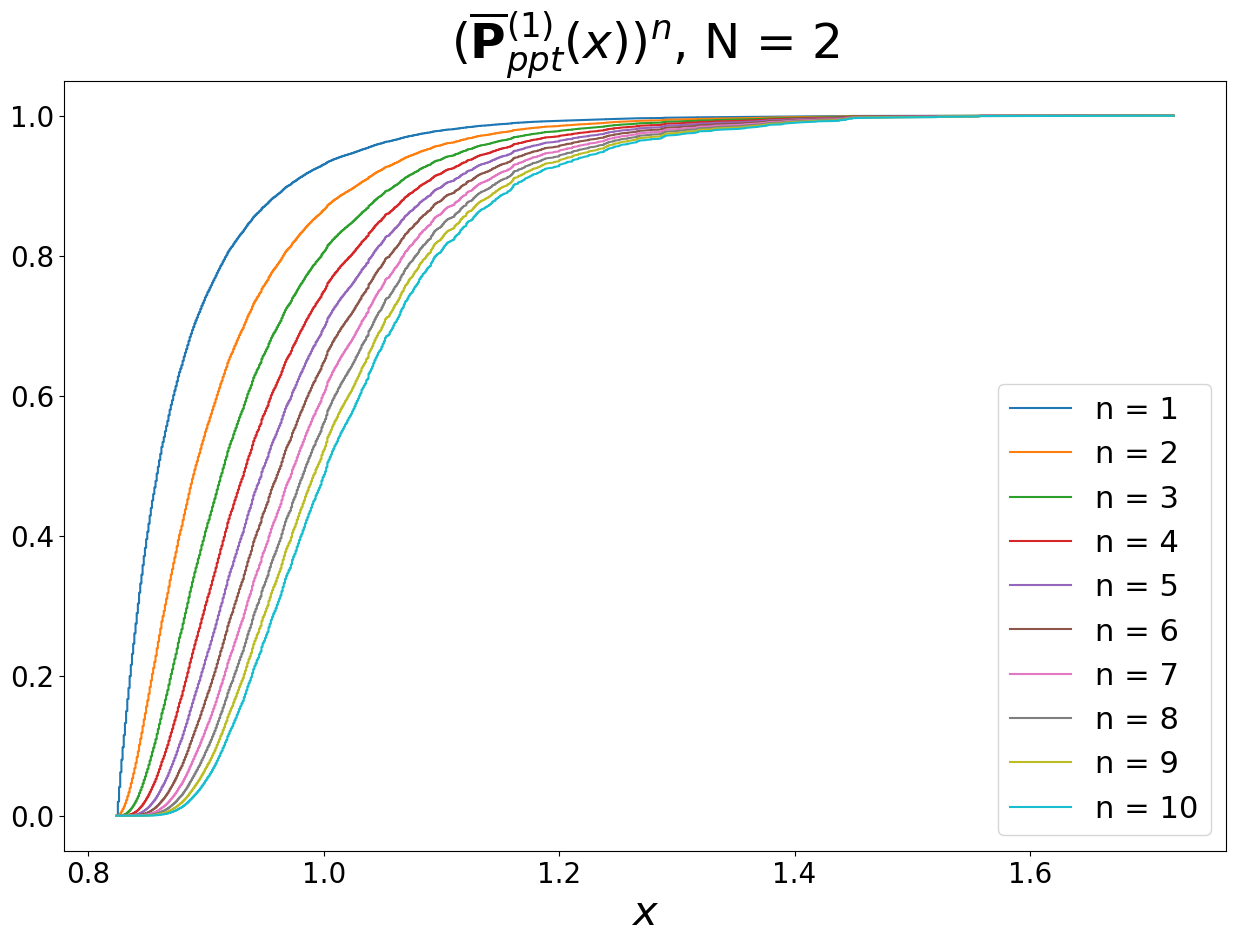}
		\caption{{Empirical cumulative distribution functions related to the PPTT distributions ${\mathbf P}^{[{\rm loc},n](k)}_{ppt}(x)$ of a collection of $n$ qubit systems, on which uniform and local random Markov noise acts (see Eq.~\eqref{easyrecursive1N}). The generator of the noise is ${\cal{L}}_{H,K}^{(k,1)}$, with $k=1000$. Plots are reported as $n$ varies from $1$ to $10$.}}
		\label{fig:cdf_n}
	\end{figure}
	The results of this analysis are summarized in Fig.~\ref{fig:cdf_n_median_and_fit}.
	 By inspecting the behaviour of the data points, a good ansatz for the fit function appears to be the following power-law scaling 
	 \begin{eqnarray}\label{fitcurveXn}
	 X_{n}\simeq \Theta_{1} n^{\Theta_{2}}\;.
	 \end{eqnarray}  By using the Minitab software, we find the values of the coefficients $\Theta_{1}$ and $\Theta_{2}$, reported in Tab.~\ref{Tab:parameter_estimates_x_n}, and the fit function together with the curves obtained considering a 95\% confidence interval for the fit parameters, reported in figure. 
	\begin{figure}[H]
		{\includegraphics[scale=0.285]{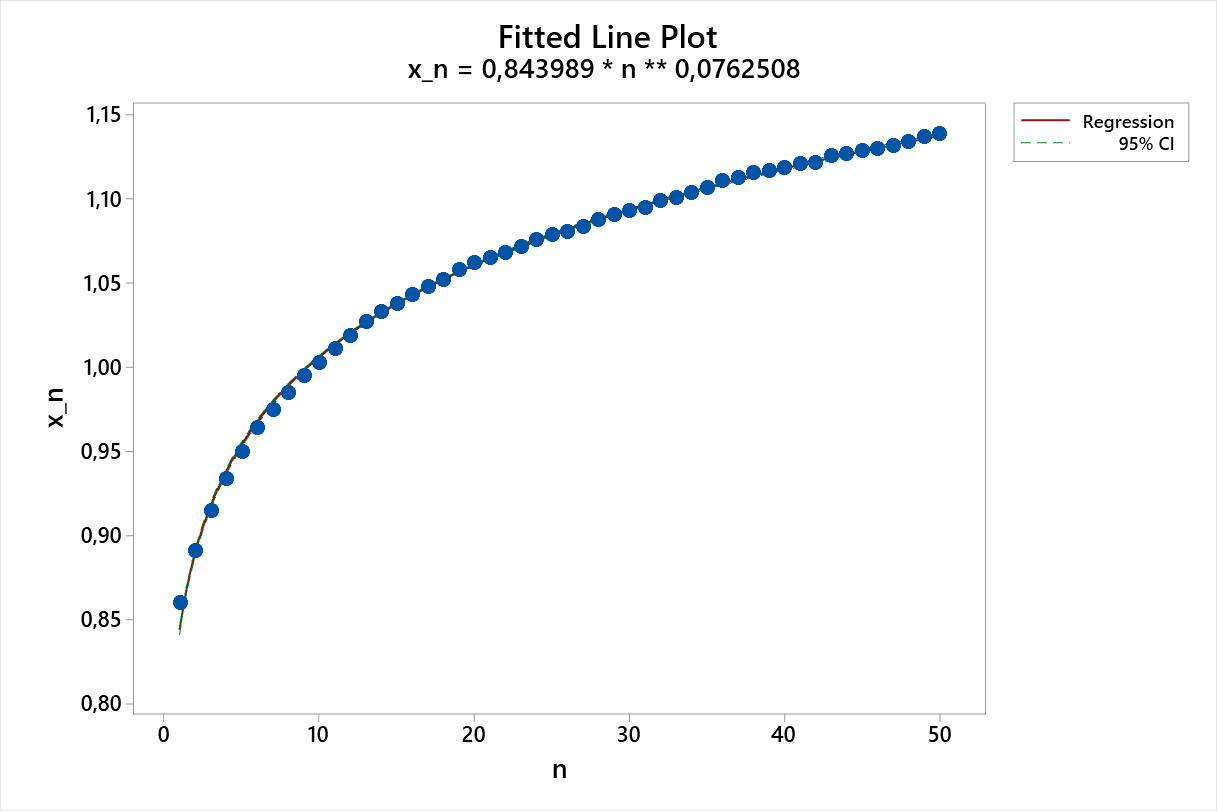}}
		\caption{ Scatterplot of points $X_{n}$ defined in Eq.~(\ref{lastlast}) 
		corresponding to the median of the PPTT distributions ${\mathbf P}^{[{\rm loc},n](k)}_{ppt}(x)$ obtained for a quantum memory of $n$ qubit systems as a function of the number $n$ of systems involved. The generator of the uniform and local random Markov noise is ${\cal{L}}_{H,K}^{(k,1)}$, with $k=1000$.
		{Dashed green lines represent the fit-curve~(\ref{fitcurveXn}) obtained considering a 95\% CI for the fit parameters (See Table \ref{Tab:parameter_estimates_x_n}).}  }
		\label{fig:cdf_n_median_and_fit}
	\end{figure}
	
	\begin{table}[H]
		\centering
		\begin{tabular}
			{cccc}
			\toprule
			Param & Estimate & SE Estimate & 95\% CI \\
			\midrule
			$\Theta_{1}$ & 0.844 & 0.001 & 	(0.841; 0.847) \\
			$\Theta_{2}$ & 0.0762 & 0.0005 & (0.0753; 0.0772) \\
			\bottomrule
		\end{tabular}
		\caption{{Parameter estimates for the  median $x_n$ of the PPTT distributions ${\mathbf P}_{ppt}(x)$ for a quantum memory of $n$ qubits $(N=2)$.}}
		\label{Tab:parameter_estimates_x_n}
	\end{table}

}}

	\section{Statistical analysis}\label{sec:stats} 
	\begin{figure*}[t]
	\subfloat[][]
	{\includegraphics[scale=0.22]{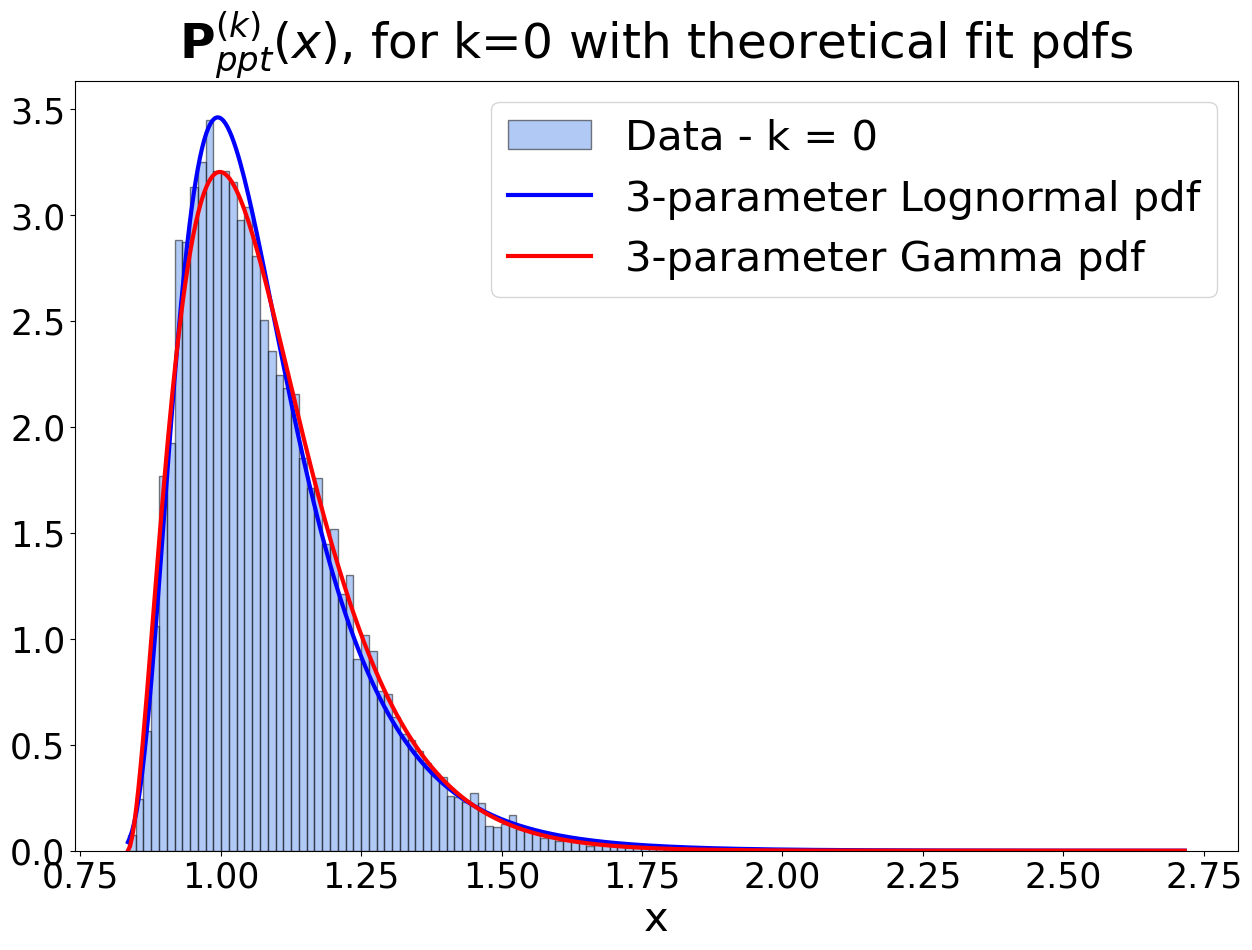}\label{fig:theor_hist_k=0}}\qquad
	\subfloat[][]
	{\includegraphics[scale=0.22]{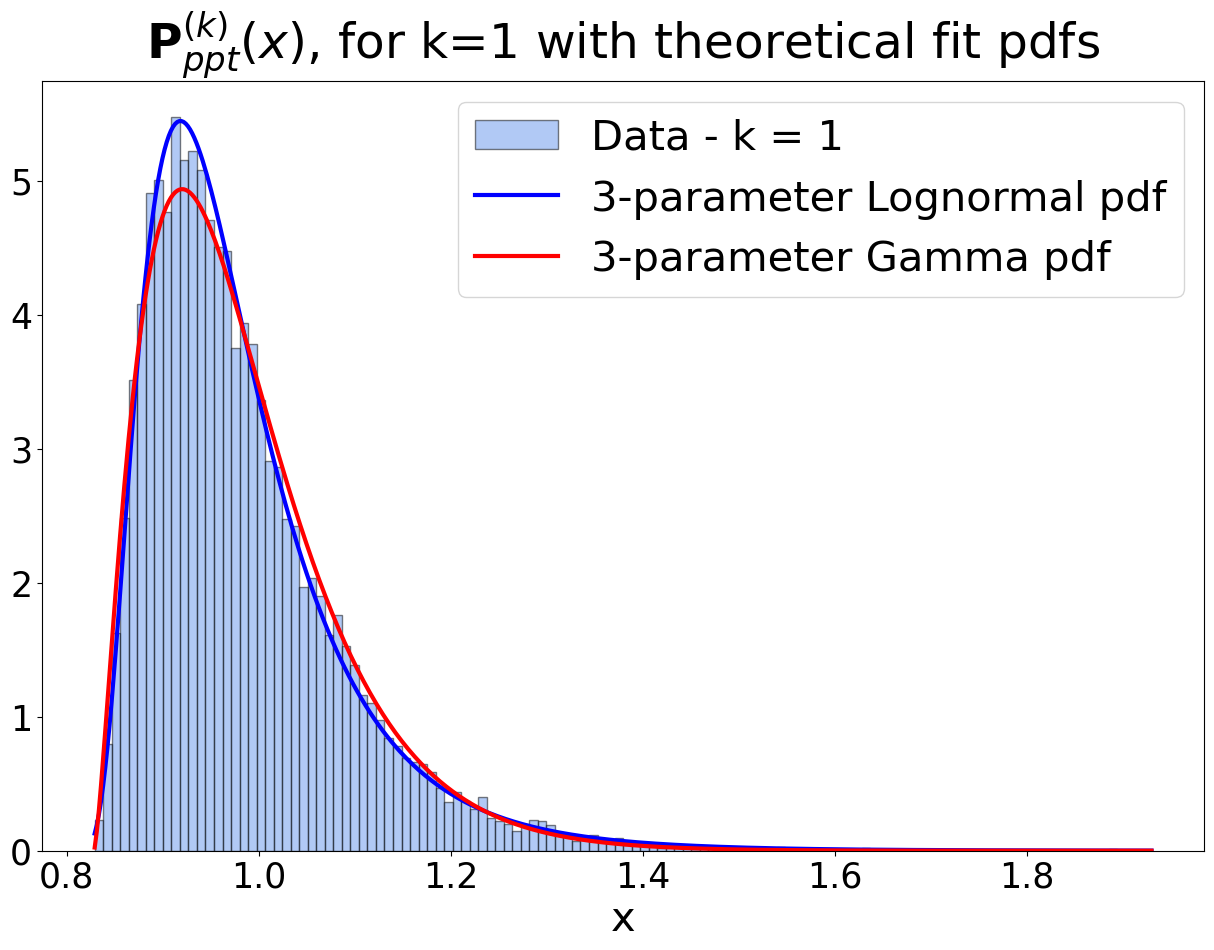}\label{fig:theor_hist_k=1}} \\
	\subfloat[][]
	{\includegraphics[scale=0.22]{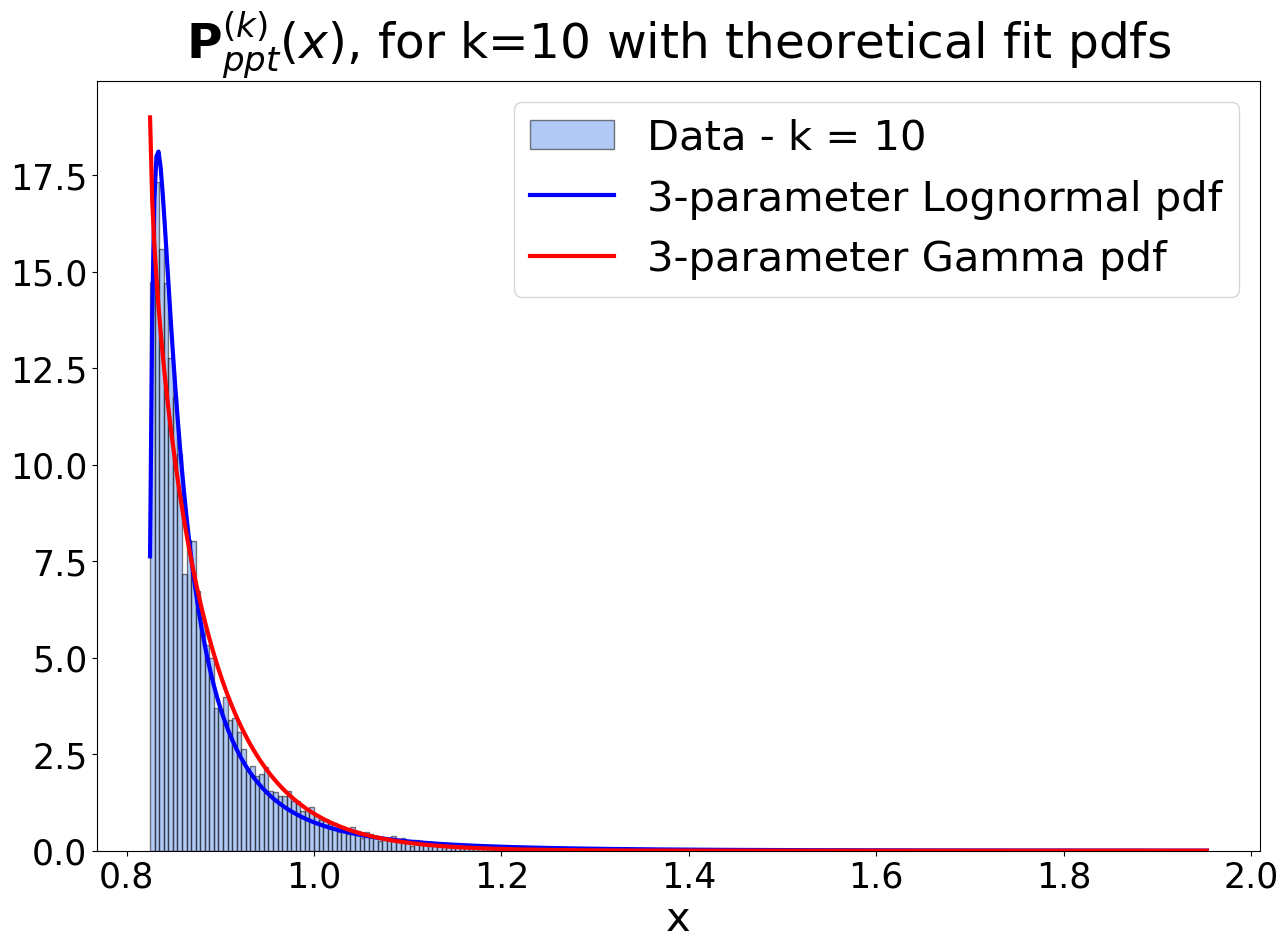}\label{fig:theor_hist_k=10}} \qquad
	\subfloat[][]
	{\includegraphics[scale=0.22]{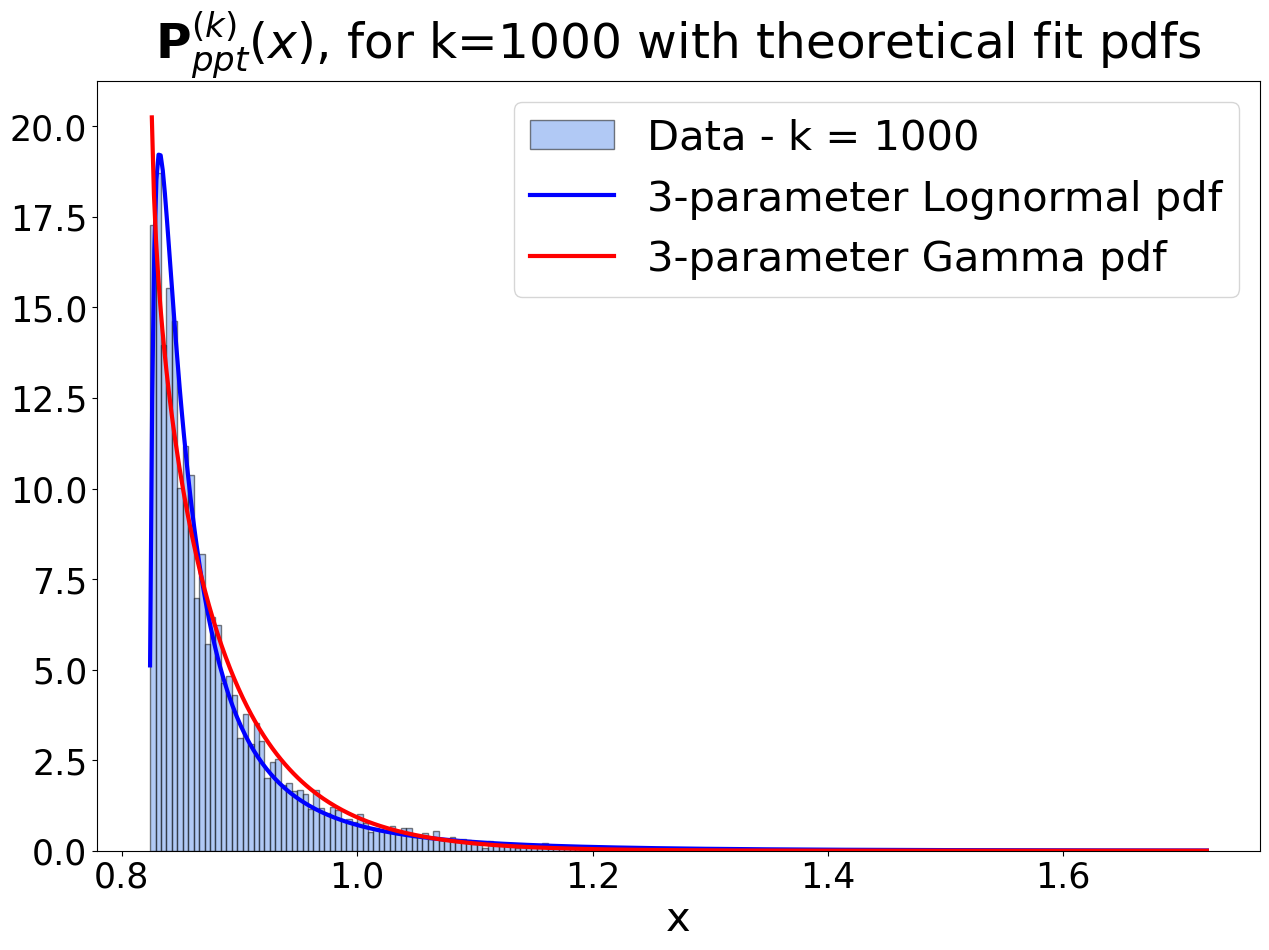}\label{fig:theor_hist_k=1000}}
	\caption{Distribution {{${\mathbf P}^{(k)}_{ppt}(x)$  of the qubit $(N=2)$ case}} for $k=0,1,10,1000$, with theoretical fit pdfs overlapped. Fit parameters reported in Table \ref{Tab: ML_estimates_param}.}
	{\label{fig:Fit_distributions_overlap}}
	\end{figure*}
In this final section we investigate the properties of the histograms derived in Sec.~\ref{sect:ESTs_distrib} obtaining fits of the probability distribution functions and studying the relations that link their characteristic times, such as mean time, median time, and minimum time to the Lindbladian parameters. Without loss of generality the analysis  will focus on the rescaled distributions
{{${\mathbf P}^{(k)}_{ppt}(x)$}} where all the quantities of interests are expressed in unit of $(1/\gamma)$ (see Eq.~(\ref{eq:Lindbladian_riscalato})): under this scenario we report results for few values of~$k$. All the details about the statistical analysis are reported in Appendix~\ref{app:distributions_stat_anlysis} for the {{qubit  ($N=2$) case. Similar consideration, indeed, also apply for the qutrits ($N=3$)  case.}}

	\subsection{Distribution fitting}
		\begin{table*}[t]
		\centering
		\begin{tabular}{clcccc}
			\toprule
			k & Distribution & Location $(\nu)$ & Shape $(\beta)$ & Scale $(\sigma)$ & Threshold $(\mu)$ \\
			\midrule
			\multirow{2}*{0} & 3P Gamma & \, & 2,89677 & 0,08656 & 0,83435 \\
			& 3P Lognorm & -1,32915 & \, & 0,49121 & 0,78692 \\
			\midrule
			\multirow{2}*{1} & 3P Gamma & \, & 2,45291 & 0,06334 & 0,82813 \\
			& 3P Lognorm & -1,86269 & \, & 0,54827 & 0,80338 \\
			\midrule
			\multirow{2}*{10} & 3P Gamma & \, & 0.9322 & 0,06660 & 0,82410 \\
			& 3P Lognorm & -3.30639 & \, & 1.10641 & 0,82249 \\
			\midrule
			\multirow{2}*{1000} & 3P Gamma & \, & 0.89219 & 0,06776 & 0,82400 \\
			& 3P Lognorm & -3.35909 & \, & 1.14009 & 0,82252 \\
			\bottomrule
		\end{tabular}
		\caption{ML estimates of distribution parameters for {{${\mathbf P}^{(k)}_{ppt}(x)$
		for the qubit $(N=2)$ case}}. The values of $k$ are relative to the plots reported in Fig.~\ref{fig:Fit_distributions_overlap}}
		\label{Tab: ML_estimates_param}
		\bigbreak
		\begin{tabular}{clcccc}
			\toprule
			k & Distribution & Location $(\nu)$ & Shape $(\beta)$ & Scale $(\sigma)$ & Threshold $(\mu)$ \\
			\midrule
			\multirow{2}*{0} & 3P Gamma & \, & 8,51657 & 0,04587	&1,33097 \\
			& 3P Lognorm & -0,64667 & \, & 0,24536 & 1,18184 \\
			\midrule
			\multirow{2}*{1} & 3P Gamma & \, & 7,30206 & 0,03678 & 1,26625 \\
			& 3P Lognorm &-1,03104 & \, & 0,26581 & 1,16539 \\
			\midrule
			\multirow{2}*{10} & 3P Gamma & \, & 3,20057 & 0,04941 & 1,24257 \\
			& 3P Lognorm & -1,66817 & \, & 0,42057 & 1,1949 \\
			\midrule
			\multirow{2}*{1000} & 3P Gamma & \, & 3,3158 & 0,04966 & 1,23287 \\
			& 3P Lognorm & -1,62198	& \, & 0,41244 & 1,1827 \\
			\bottomrule
		\end{tabular}
		\caption{ML estimates of distribution parameters for {{${\mathbf P}^{(k)}_{ppt}(x)$
		for the qutrit $(N=3)$ case}}. The values of $k$ considered are the same as for $N=2$.}
		\label{Tab: ML_estimates_param_N=3}
	\end{table*}
	 For each data set associated with the sampling of the  Lindbladian $\mathcal{L}_{H,K}^{(k,1)}$ with $k$ fixed,
	 we used the Minitab software to carry out the statistical analysis, exploiting the maximum likelihood method to find the optimal parameters of the chosen model and the probability plots and goodness-of-fit tests to understand which distribution represents the best fit for the data. From this analysis, it emerges that there exist two distributions that could represent a good fit for the data, which are the 3-parameter Gamma distribution (Eq.~\eqref{Eq:3P-Gamma}) and the 3-parameter Lognormal distribution (Eq.~\eqref{Eq:3P-Lognormal}), with no elements that justify the propensity for one distribution rather than the other.
	 \begin{equation}
	 	f_{G}^{(\beta,\sigma,\mu)}(x) = \frac{(x-\mu)^{(\beta-1)}}{\sigma^{\beta}\Gamma(\beta)}\exp\Big(-\frac{(x-\mu)}{\sigma}\Big),
	 	\label{Eq:3P-Gamma}
	 \end{equation}
 	 \begin{equation}
 	  	f_{LN}^{(\sigma,\mu,\nu)}(x) = \frac{(x-\mu)^{-1}}{\sqrt{2\pi} \sigma}\exp\Big(-\frac{(\ln(x-\mu) - \nu)^2}{2\sigma^2}\Big).
 	  	\label{Eq:3P-Lognormal}
 	 \end{equation}
   	 Here $\beta$ is the shape parameter, $\sigma$ is the scale parameter, $\mu$ is the threshold value and $\nu$ is the location.
	 In Fig.~\ref{fig:Fit_distributions_overlap}  we report a comparison between these functions and the corresponding data set of {{${\mathbf P}^{(k)}_{ppt}(x)$ of the qubit $(N=2)$ case}} for some values of $k$.
	The parameters obtained from the statistical analysis, using the maximum likelihood (ML) method, are reported in the Table~\ref{Tab: ML_estimates_param}.
	
	The same analysis has then been conducted considering the data set for ${\mathbf P}^{(k)}_{ppt}(x)$ {{of the qutrit 
	$(N=3)$ case}}, obtaining the same results in terms of fit distributions, whose characteristic parameters are reported in Table \ref{Tab: ML_estimates_param_N=3}.
	
	\begin{figure*}[!htb]
		\subfloat[][]
		{\includegraphics[scale=0.215]{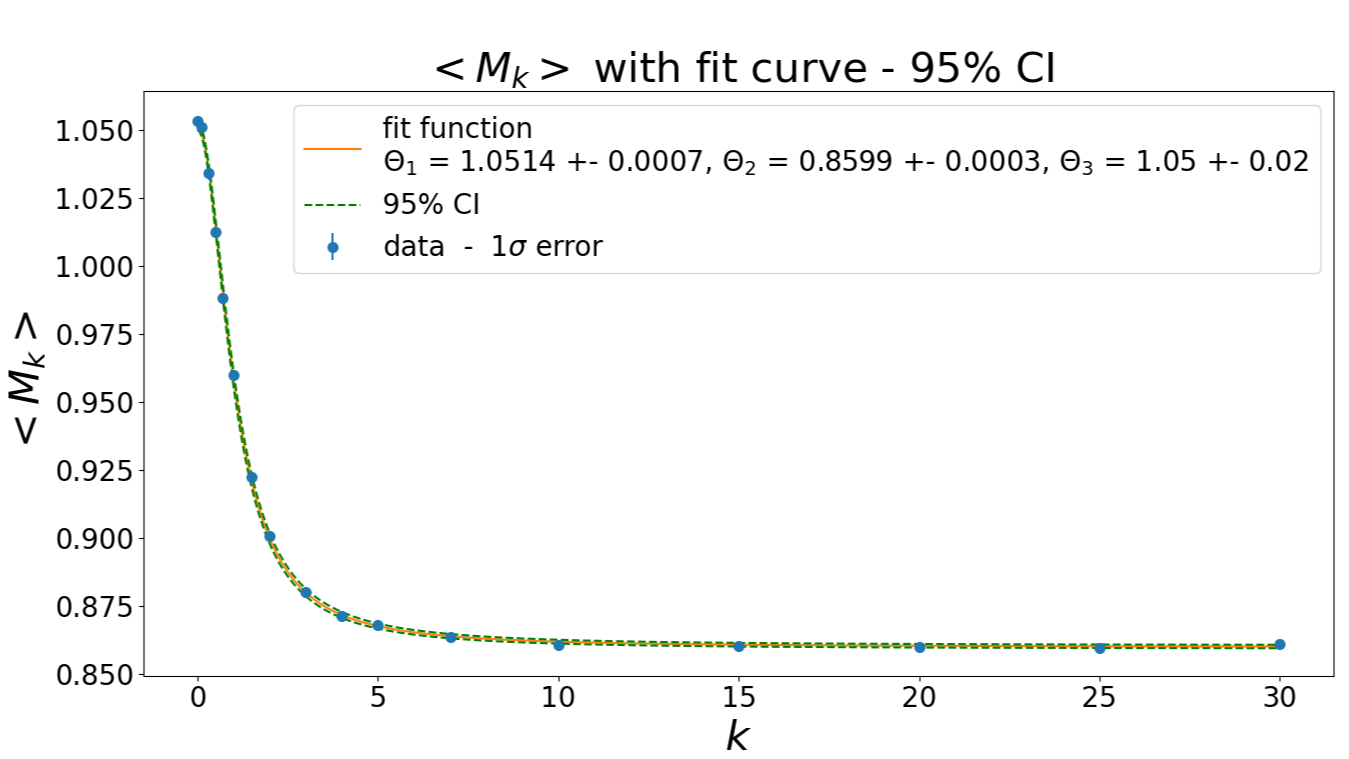}\label{fig:fit_t_ent_panel_a}}\quad
		\subfloat[][]
		{\includegraphics[scale=0.215]{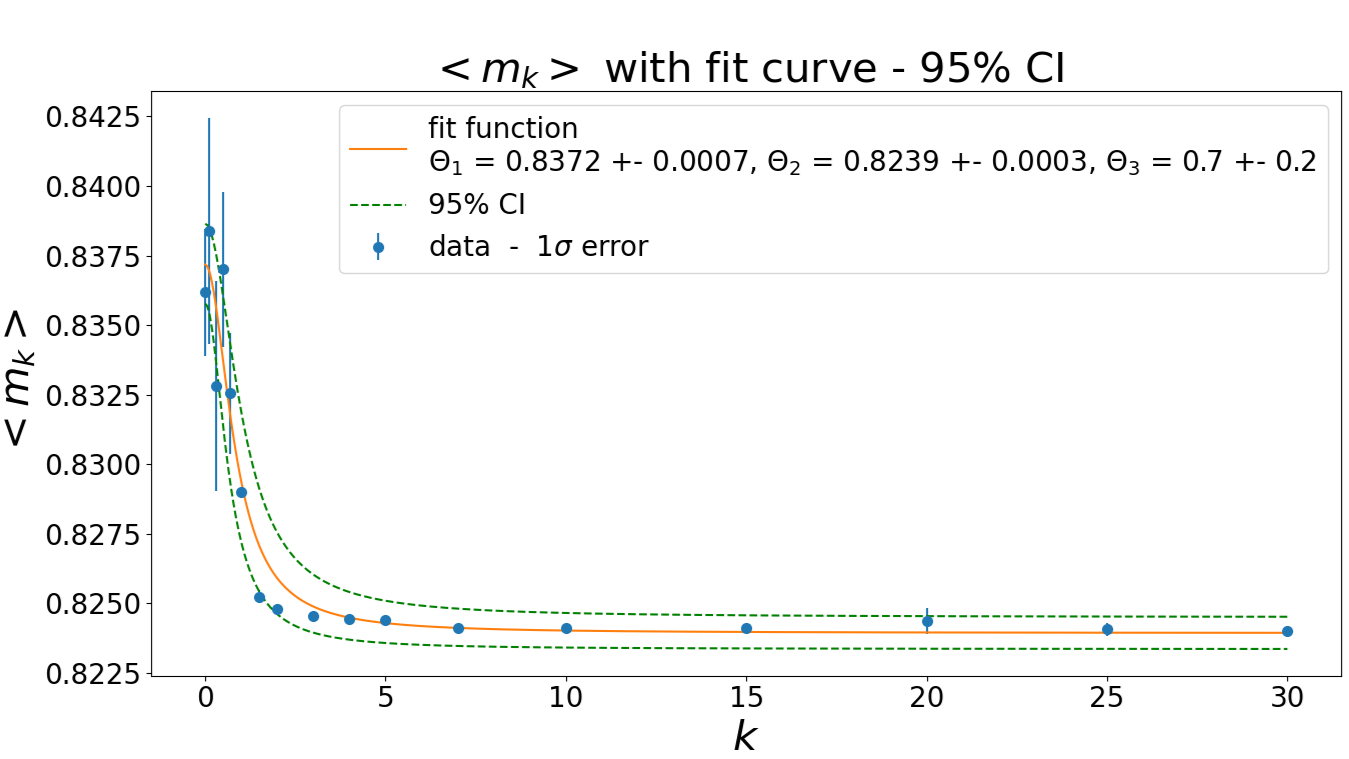}\label{fig:fit_t_ent_panel_b}}
		\caption{Panel (a): Functional dependence of the median $\langle M_k \rangle$ 
			of the distribution {{${\mathbf P}^{(k)}_{ppt}(x)$
		for the qubit $(N=2)$ case}}
			 obtained from the data of the histogram of Fig.~\ref{fig:hist_panel_a}. Dashed green lines represent the fit-curves obtained considering a 95\% CI for the fit parameters (See Table \ref{Tab:parameter_estimates_Me(tau_ent)} in Appendix \ref{app:charctEST_stat_anlysis}). Panel (b): Functional dependence of minimum time  $\langle m_k \rangle $ of the distribution  {{${\mathbf P}^{(k)}_{ppt}(x)$
		for the qubit $(N=2)$ case}} obtained from the data of the histogram of Fig.~\ref{fig:hist_panel_a}. Dashed green lines represent the fit-curves obtained considering a 95\% CI for the fit parameters (See Table \ref{Tab:parameter_estimates_min(tau_ent)} in Appendix \ref{app:charctEST_stat_anlysis}).}
		\label{fig:fit_t_ent}
		\bigbreak
		\subfloat[][]
		{\includegraphics[scale=0.215]{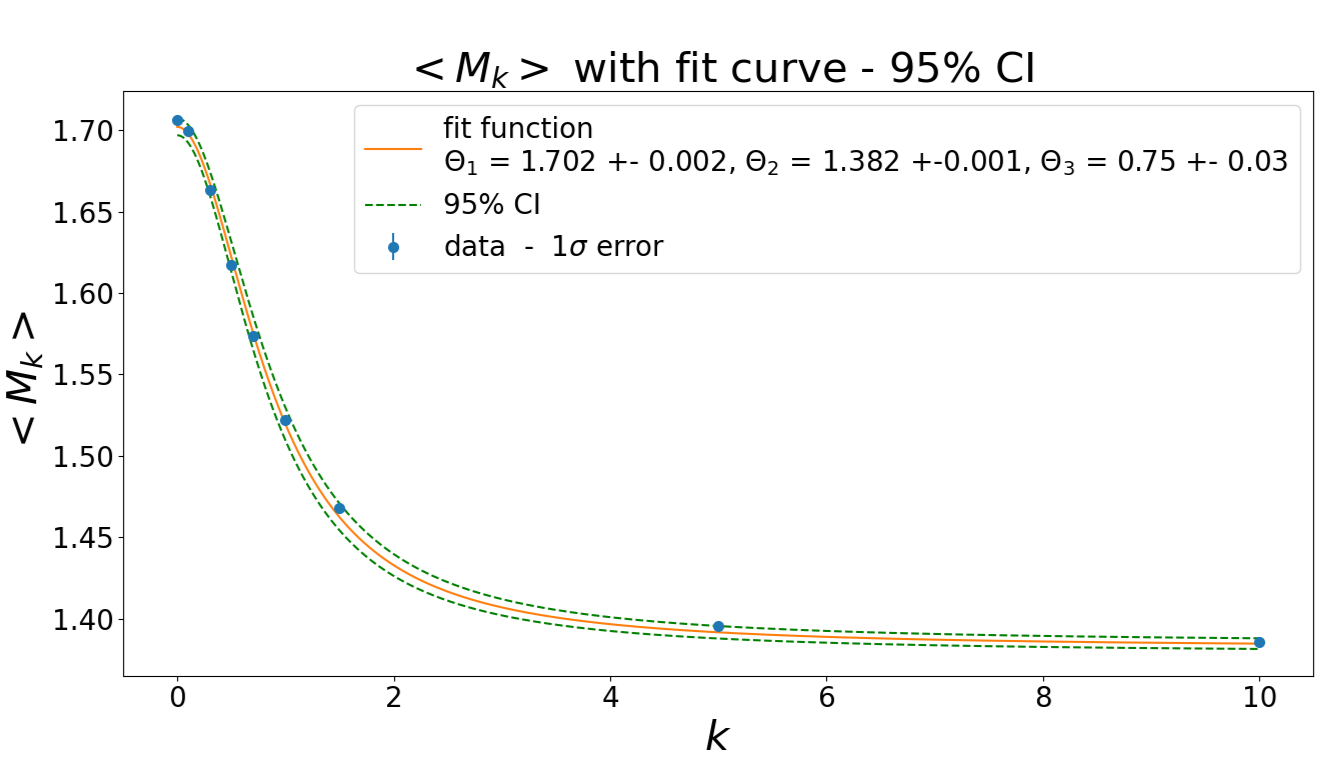}\label{fig:fit_t_pptN3_panel_a}}\quad
		\subfloat[][]
		{\includegraphics[scale=0.215]{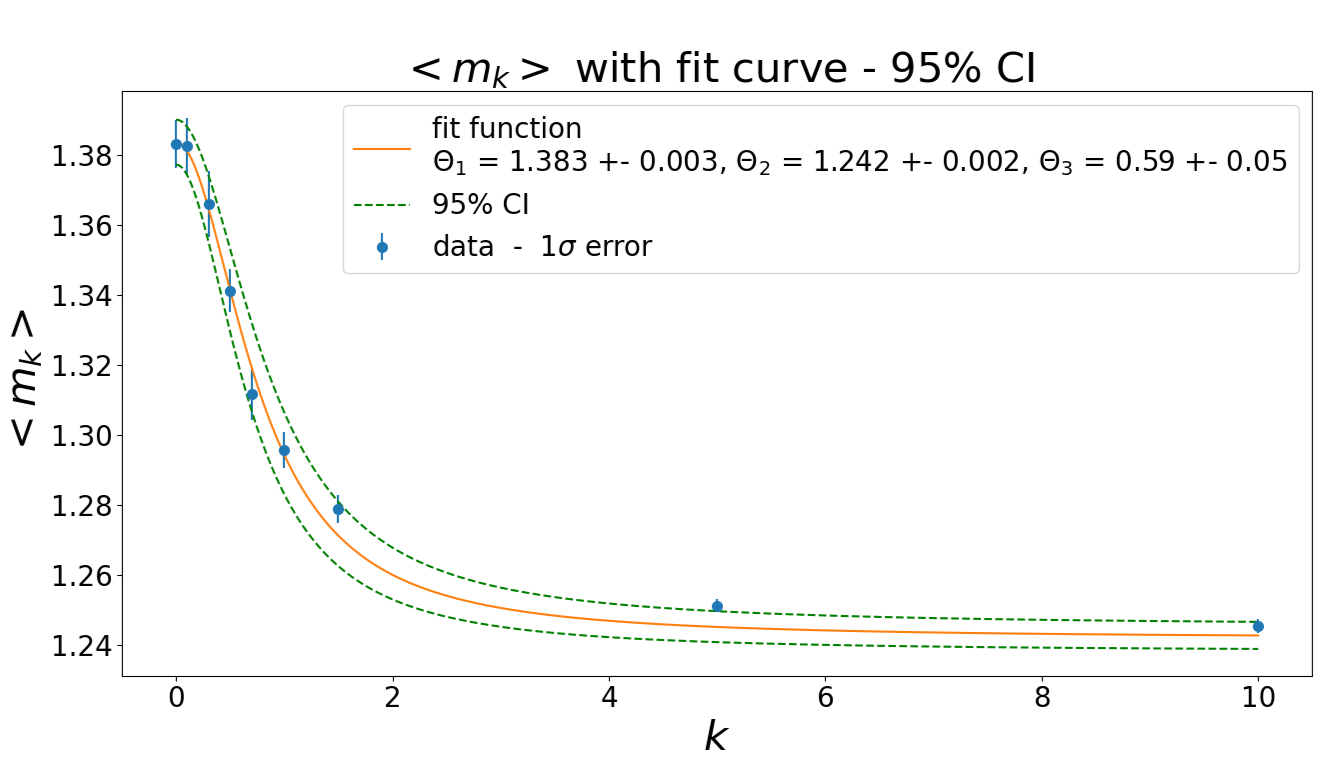}\label{fig:fit_t_pptN3_panel_b}}
		\caption{{{Same plot as in Fig.~\ref{fig:fit_t_ent} for the distribution ${\mathbf P}^{(k)}_{ppt}(x)$
		for the qutrit $(N=3)$ case. Here the reference data are taken from Fig.~\ref{fig:histN3_panel_a} while
		the fit parameters are defined in Table \ref{Tab:param_estim_m_k_N=3} of Appendix~\ref{app:charctEST_stat_anlysis}.
		}}
			}
		\label{fig:fit_t_pptN3}
	\end{figure*}
	{{
	\subsubsection{Analysis of characteristic PPTTs}{\label{subsec:analysis_char_EST}}
	
	We can now analyze the PPTT distributions ${\mathbf P}^{(k)}_{ppt}(x)$}} in more detail studying how their characteristic times, such as mean time 
	$\mu_k$, median time ${M}_k$, and minimum time $m_k$ which are the quantities of interest in most cases, vary with respect to the parameter $k$~\cite{NOTA3}. 
	 First and foremost one expects that all the quantities will be expressed by functions   which are even in $k$, as positive and negative values of $k$ formally corresponds to positive and negative values of $\alpha$ which will contribute equally to the dynamics when
	 sampling  $\hat{H}$ in $\mbox{GUE($N$)}$.
	 Moreover, from the limit behaviour of the distributions {{${\mathbf P}^{(k)}_{ppt}(x)$}}, which can be investigated from the data, it emerges that all 
	 the searched functions $\mu_k$, ${M}_k$, and $m_k$
	 must tend to a constant value both for $k$ approaching zero and for $k$ approaching infinity. Based on these considerations, the ansatz we have made is that they will be expressed  as ratios $f_{\vec{\theta}}(k)$ of second-order polynomials in $k$, with three parameters
	 $\vec{\theta}:=(\theta_1,\theta_2,\theta_3)$  to be determined, i.e.
	 \begin{equation}
		f_{\vec{\theta}}(k) = \frac{\theta'_{1} + \theta_{2}k^2}{\theta_{3} + k^2}\;, 
		\label{eq:fit_function}
	\end{equation}
where $\theta_{1}'= \theta_{1}\theta_{3}$, and
	\begin{equation}
			{\theta_{2}} = \lim_{k  \rightarrow \infty} f_{\vec{\theta}}(k)\;, \qquad 
			{\theta_{1}} =f_{\vec{\theta}}(0)\;.
	\end{equation}
	As discussed in Appendix \ref{app:charctEST_stat_anlysis} the special  choice of making $\theta_1'$ proportional to $\theta_3$ is made to reduce the parameter estimation error and the correlation otherwise present between the $\theta_{1}$ and $\theta_{3}$.

 In the following, we present 
 the results obtained in using the functions  $f_{\vec{\theta}}(k)$ as estimators
 for the characteristic times of the problem. In particular, in order to associate an error to data points, the characteristic times reported in Figs.~\ref{fig:fit_t_ent}, \ref{fig:fit_t_pptN3} have been obtained from the original sample using the Bootstrap method. Hence, they must be understood as average values and therefore they are indicated as  $\langle M_{k}\rangle, \langle \mu_{k}\rangle, \langle m_{k}\rangle$. All the details can be found in the Appendix \ref{app:charctEST_stat_anlysis}. As can be seen from Figs.~\ref{fig:fit_t_ent_panel_a}, \ref{fig:fit_t_pptN3_panel_a}, the fit function~(\ref{eq:fit_function}) 
  well describes the functional dependence of the median time  $\langle M_k \rangle$ from $k$.
  Very similar results, in terms of fit precision, have been obtained also for   the mean time $\langle \mu_k \rangle$ (data not reported). 
   On the contrary for the minimum time $\langle m_k \rangle$ the data points are characterized by much larger errors than the ones associated to $\langle M_k \rangle $, when one considers small $k$ values, and small errors that appear for large $k$ values (see Fig.~\ref{fig:fit_t_ent_panel_b}, \ref{fig:fit_t_pptN3_panel_b}). This trend can be justified by remembering that the minimum time strongly depends on the number of iterations chosen in the numerical simulation: for the qubit case, when $k$ assumes small values, the {{PPTT}}  distributions are characterized by a smooth right-skewed form, therefore the minimum time can be strongly dependent from the sampling, causing the associated errors to be larger. When $k$ increases, instead, these distributions tend to a strongly peaked limit distribution, with the peak in correspondence with the minimum time, therefore such times tend to be estimated more precisely. Similar considerations also apply to the qutrit case.

	As a final remark we notice that from our numerical data it  emerges that in all the cases the optimal fitting parameters 
	$\theta_1$ is always greater than the associated $\theta_3$, implying that, in agreement with the observation of~\cite{Gatto} 
	 the characteristic times that are obtained from {{the PPTT}} distributions, are all monotone functions, decreasing in $k$.

	\section{Conclusions}\label{sect:conclusion}
{{	The current investigation delves into the distributions of positive partial transpose times for quantum system subjected to the influence of  an ensemble of Markovian noises. Our primary focus lies in examining the correlation between the characteristics of these distributions and the parameters governing the relative intensities of the Hamiltonian and dissipative components within the Lindbladian super-operator.
Particularly noteworthy is our characterization of the limit distribution observed in the weak-coupling regime, where the predominance of the unitary Hamiltonian term is pronounced. This facet of our study draws inspiration from the findings in Ref.~\cite{Gatto}, albeit in our case, we explore a broader spectrum of noise models under the same limiting circumstances. 
As a case of special interest we also consider the case where the system of interest is composed by a collection of $n$ elements (many-body configuration) subject to local, independent random evolutions: a numerical analysis of this configuration shows that the charactistic time at which the system has probability 1/2 of reaching PPTT starting from a maximally entangled input configuration, scales with a power law w.r.t. the system size. 
While our investigation is centered on the seemingly straightforward yet non-trivial instances of qubit and  qutrit systems, we posit that this research could serve as a stepping stone for broader generalizations. Specifically, we contemplate the extension of our approach to more complex systems, exploring whether and how the results observed in these cases might extrapolate. Additionally, we foresee the potential to glean insights into positive partial transpose times when introducing new symmetries or imposing different constraints on the noise within the system of interest.}}
\\		
		
		NC and VG 
acknowledge financial support by MUR (Ministero dell’ Universit\`a e della Ricerca) through the PNRR MUR project PE0000023-NQSTI.
GDP has been supported by the HPC Italian National Centre for HPC, Big Data and Quantum Computing - Proposal code CN00000013 - CUP J33C22001170001 and by the Italian Extended Partnership PE01 - FAIR Future Artificial Intelligence Research - Proposal code PE00000013 - CUP J33C22002830006 under the MUR National Recovery and Resilience Plan funded by the European Union - NextGenerationEU.
Funded by the European Union - NextGenerationEU under the National Recovery and Resilience Plan (PNRR) - Mission 4 Education and research - Component 2 From research to business - Investment 1.1 Notice Prin 2022 - DD N. 104 del 2/2/2022, from title ``understanding the LEarning process of QUantum Neural networks (LeQun)'', proposal code 2022WHZ5XH – CUP J53D23003890006.
GDP is a member of the ``Gruppo Nazionale per la Fisica Matematica (GNFM)'' of the ``Istituto Nazionale di Alta Matematica ``Francesco Severi'' (INdAM)''.

\appendix
	
	\bibliographystyle{apsrev4-1} 
	\bibliography{xampl} 

\begin{thebibliography}{20}
\bibitem{NOTA2}  In our analysis we shall assume $\alpha, \gamma$  to have 
  the dimension of a frequency ($\hbar = 1$) in order to make $\hat{H}$ and $K$ dimensionless.
\bibitem{NOTA3} It is worth remarking that from (\ref{eq:scaling_law_beta}) it follows 
that the corresponding mean $\mu_{\alpha, \gamma}$, median $M_{\alpha, \gamma}$, and minimum 
$m_{\alpha, \gamma}$ times of the distribution $P^{(\alpha,\gamma)}_{ent}(\tau)$
are linked with $\mu_k$, $M_k$, and $m_k$ of ${\mathbf P}^{(k)}_{ent}(x)$ via simple riscalings
$\mu_{\alpha, \gamma}  = \frac{1}{\gamma} \mu_{\alpha/\gamma}$, 
$M_{\alpha, \gamma}  = \frac{1}{\gamma} M_{\alpha/\gamma}$, and 
$m_{\alpha, \gamma}  = \frac{1}{\gamma} m_{\alpha/\gamma}$.
\end{thebibliography}

	\appendix
{{\section{Necessary and sufficient condition for PPT} \label{appePPT}
Here we give an explicit prove that a generic  $\Phi$ LCPT superoperator is PPT if and only if the Choi-Jamiołkowski state~\cite{Choi-Jam} $\hat{\rho}_{SS'}^{\Phi}:=\Phi\otimes {\rm{Id}}_{S'}(|{\Psi_{\max}}\rangle_{SS'}\langle {\Psi_{\max}}|)$ 
is PPT, i.e. 
\begin{eqnarray} \label{equivalence}
	\Phi \in  \text{PPT} \quad \Leftrightarrow \quad \left(\hat{\rho}_{SS'}^{\Phi}\right)^{T_{S'}}
	 \geq 0\;,  \label{impo22211} 
	\end{eqnarray}
	or equivalently 
	\begin{eqnarray} \label{equivalence1}
	\Phi \in  \text{PPT} \quad \Leftrightarrow \quad \left(\hat{\rho}_{SS'}^{\Phi}\right)^{T_{S}}
	 \geq 0\;,  \label{impo222111} 
	\end{eqnarray}
	with $T_{S}$ and $T_{S'}$ representing the partial transpositions w.r.t. to the system $S$ and $S'$ respectively. 
	The direct implication in Eq.~(\ref{equivalence}) is a consequence of the definitions. To prove the reverse observe that given $|\Psi\rangle_{SS'}$ a generic pure state of
$SS'$, there exists a local operator $\hat{X}_{S'}$  on $S'$ such that 
\begin{eqnarray} 
|\Psi\rangle_{SS'} = \hat{X}_{S'} |{\Psi_{\max}}\rangle_{SS'}\;,
\end{eqnarray} 
so that 
\begin{eqnarray} 
 \hat{X}_{S'}  \hat{\rho}_{SS'}^{\Phi} \hat{X}^\dag_{S'} = \Phi\otimes {\rm{Id}}_{S'}(|{\Psi}\rangle_{SS'}\langle {\Psi}|).
\end{eqnarray} 
If hence  $\hat{\rho}_{SS'}^{\Phi}$ is PPT we can write 
\begin{eqnarray} 
&& \left( \Phi\otimes {\rm{Id}}_{S'}(|{\Psi}\rangle_{SS'}\langle {\Psi}|)\right)^{T_{S}} \nonumber \\ 
&&\qquad = \hat{X}_{S'}   \left(  \hat{\rho}_{SS'}^{\Phi} \right)^{T_{S}} \hat{X}^\dag_{S'} \geq 0\;.
\end{eqnarray} 
The extension to a generic mixed input state follows finally by linearity.

 }}
	\section{Statistical Analysis of EST distributions}{\label{app:distributions_stat_anlysis}}
	The distributions of entanglement survival time have been studied from a statistical point of view in order to find a theoretical distribution that could represent a discrete fit for the data. In particular, we used the the Minitab software, in which it is possible to test up to 14 theoretical distributions.
	For each distribution tested, the probability plot is shown and Minitab calculates Anderson Darling statistics (AD)	and the associated p-value. In the probability plot, the input data is plotted with respect to the percentage of values in the sample that is less than or equal to the considered value, so that they roughly follow a straight line if they come from the chosen distribution. In particular, these graphs show a central straight line, which corresponds to the expected theoretical percentiles, calculated considering the theoretical distribution, evaluated in correspondence with the parameters obtained with the maximum likelihood method. Lines at a 95\% confidence interval for the model parameters are also added. Thus, deviations from the central line represent deviations from the considered distribution. Associated with these graphs is the AD coefficient: this value gives a measure of the distance between the numerically found distribution and the theoretical one and it is calculated considering the cumulative distribution. Being a distance measure, smaller values of this parameter indicate that the data better follow the theoretical distribution. Consequently, during the analysis, among the considered fit distributions, those with a lower AD value are considered better than the others. Furthermore, the AD value allows the calculation of the p-value, which represents a probability that measures the evidence against the null hypothesis. For this test, the null hypothesis is that the data follow the tested distribution. Consequently, p-values less than a certain threshold (the threshold is fixed at $\bar{\alpha} = 0.05$) indicate that the data do not follow that distribution, since the null hypothesis is rejected. For certain classes of distributions reported in the analysis, Minitab also provides the p-value associated with the likelihood ratio test. This test is used to understand if the addition of a parameter substantially changes the fit and the null hypothesis at its base is that the distribution that best fits the data is the one with the fewest parameters, i.e.,~the simplest model. In this case, therefore, having a significance level of $\bar{\alpha}=0.05$, with low p-values it can be concluded that the data follow the distribution with a greater number of parameters. By looking at these indicators it is possible to get an idea of which distribution is the best.
	This analysis was made both for the qubit and the qutrit case for a subset of the investigated $k$ values, in the case of constant $\gamma$ and variable~$\alpha$. Below are the considerations made to arrive at the chosen fit distribution, in the two-qubit case, for the value $k=1$, which is the intermediate case par excellence, where both contributions are present with the same strength. For all the other cases the same logic was followed, reaching the same conclusions. An initial evaluation of the probability plots shows that the only distributions that can represent a good fit for the data are the 3-parameter Gamma distribution and the 3-parameter Lognormal distribution. This can be deduced from the good graphic agreement between the experimental points and the theoretical line and from the AD values, significantly lower than those of the other distributions. Here is reported the summary plot where these distributions are compared with the 2-parameter ones.
	\begin{figure}[H]
		\centering
		\includegraphics[scale=0.3]{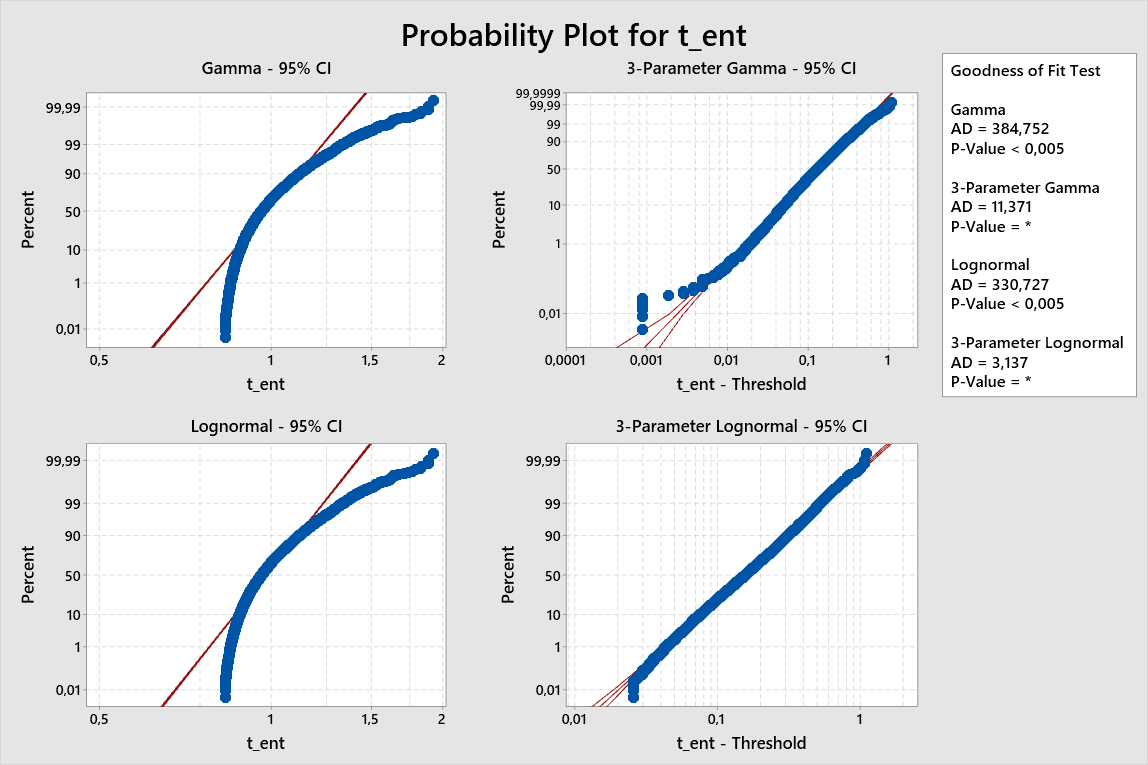}
		\caption{Distribution identification for $N=2$, $k=1$; probability plot and goodness-of-fit test.}
		\label{fig:gamma_lognormal_fit_k=1}
	\end{figure}
	It should be noted that for the 3-parameters distributions it is not possible to define a p-value, but it is possible to compare the p-values associated with the likelihood ratio criterion. Below is the table of the goodness-of-fit test obtained when considering the Gamma and Lognormal distributions
	\begin{table}[H]
		\centering
		\begin{tabular}{lccc}
			\toprule
			Distribution & AD & P & LRT P \\
			\midrule
			Gamma & 250,107 & $<$0,005 & \, \\
			3-Parameter Gamma & 5,704 & * & $<$0,001 \\
			Lognormal & 197,695 & $<$0,005 & \, \\
			3-Parameter Lognormal & 7,802 & * & $<$0,001 \\
			\bottomrule
		\end{tabular}
		\caption{Goodness-of-fit test: distribution identification for $N=2$, $k=0$.}
		\label{Tab: Goodness-of-fit_k=0}
	\end{table}
	This table shows that the p-value associated with the likelihood ratio test is less than the significance level for both distributions. This means that the addition of a parameter significantly improves the fit in both cases, but does not give information on which of the two distributions best fits the experimental data. To obtain further information in this sense, we have compared the log-likelihood values for both distributions, using as parameters those estimated by Minitab with the maximum likelihood estimation method
	\begin{table}[H]
		\centering
		\begin{tabular}{lc}
			\toprule
			Distribution & log-likelihood \\
			\midrule
			3-Parameter Gamma & 12429,3 \\
			3-Parameter Lognormal & 12422,1\\
			\bottomrule
		\end{tabular}
		\caption{Log-likelihood associated with the 3-Parameter Gamma and 3-Parameter Lognormal distributions for $N=2$, $k=0$, computed using ML estimates of distribution parameters.}
		\label{Tab: log_likelihood for k=0}
	\end{table}
	As can be seen, the log-likelihood values of the two distributions are very similar and their difference would not justify the propensity for one fit distribution rather than another. Therefore, it is concluded that both theoretical distributions can represent a good fit for the experimental data.
	
	\section{Analysis of Characteristics ESTs and PPTTs}{\label{app:charctEST_stat_anlysis}}
	  To determine the optimal values of the parameters 
	  $\theta_{1}',\theta_{2},\theta_{3}$ which define the  fit function $f_{\vec{\theta}}(k)$ of Eq.~(\ref{eq:fit_function}) we 
	  minimize 
	   the sum of squares of the residual error (SSE) through an iterative procedure which has to be terminated when a convergence criterion is satisfied.
	\begin{figure*}[!htb]
		\centering
		\includegraphics[scale=0.3]{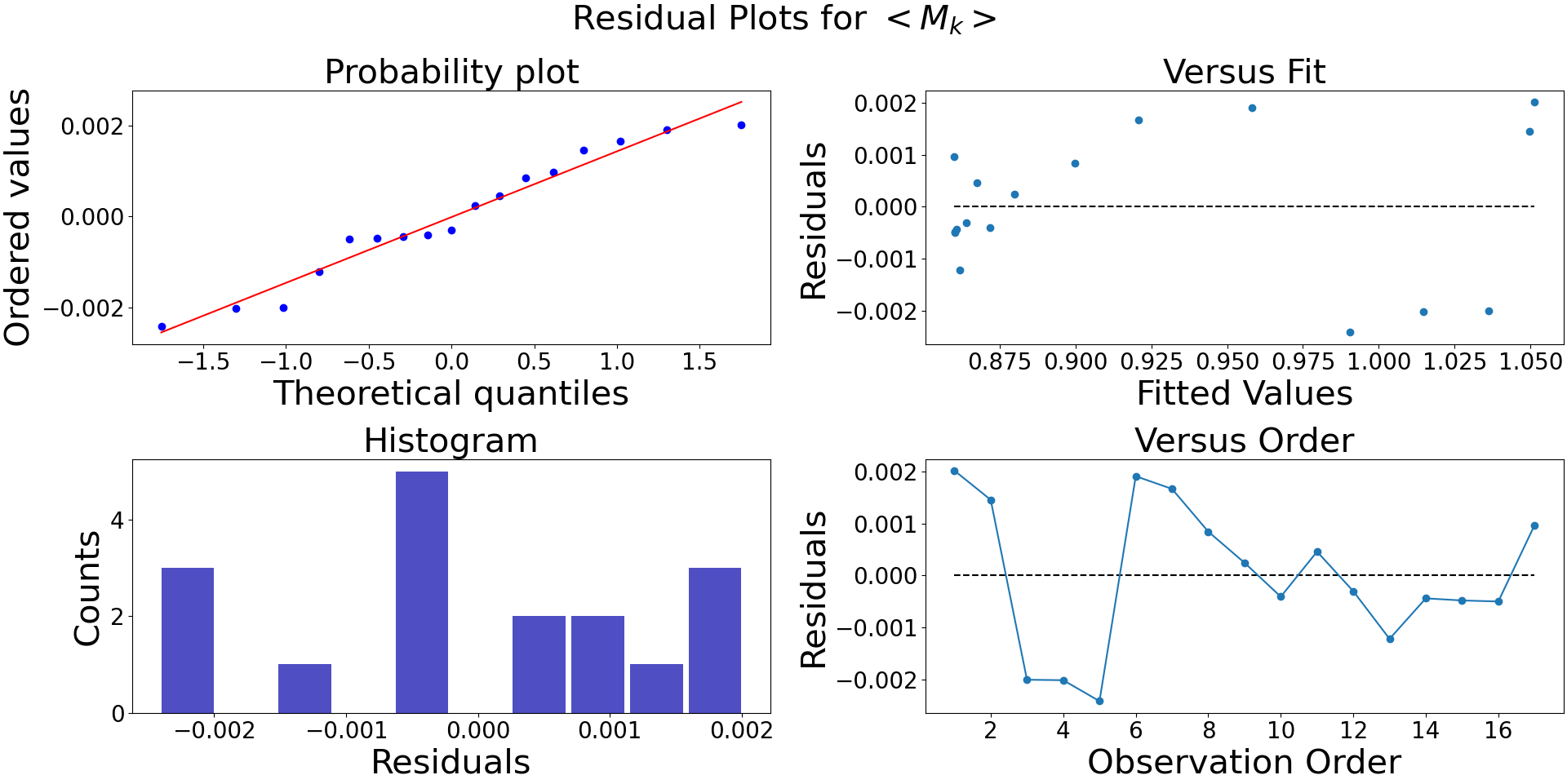}
		\caption{Residual plots for $M_k$ in the two-qubit case $(N=2)$ (maximum value of $k$ being $30$).}
		\label{fig:t_ent_Me_residual_plots}
	\end{figure*}
	To perform the analysis we used the Minitab software which exploits the Levenberg-Marquard algorithm, choosing 200 as the maximum number of iterations that the algorithm can use to achieve convergence and $0.00001$ as the convergence tolerance.	Once the algorithm converged, it returns the parameter estimates, with a $95\%$ confidence interval, and the correlation matrix for them. Thanks to this matrix it was possible to see that, in all the considered cases, if one chooses the first parameter as $\theta_{1}'= \theta_{1}$, a strong correlation between $\theta_{1}'$ and $\theta_{3}$ appears that also reflects in a huge uncertainty in the parameter estimates. This is due to the fact that, when considering the function's values for $k=0$, one would get   
	$f_{\vec{\theta}}(0) = \theta_1/\theta_3$.
	To avoid this  it was chosen $\theta_{1}'= \theta_{1}\theta_{3}$: this choice eliminates partially the correlation and also improves the estimation of these parameters. 
	To associate an error to data points we have used the Bootstrap method and from the original sample with $20000$ iterations, we have created $100$ new samples, each with the same number of elements as the original one. Therefore, the points that are represented in the graphs \ref{fig:fit_t_ent}, \ref{fig:fit_t_pptN3} must be understood as average values over $100$ points and the error associated with them is equal to $1\,\sigma$. In order to find the parameter estimates with the best precision possible, we have used in the analysis all the values of $k$, up to $k=1000$, while, for greater clarity of visualization, a close-up of the original plots, up to $k=30$, is reported here for the two-qubit case. Finally, residual analysis is considered in order to assess the goodness of fit. In particular, we considered the normal probability plot of residuals, the plot of the residuals versus fitted values, the histograms of the residuals, and the plot of the residuals versus order. These plots allow us to understand if the assumptions that are at the basis of the regression analysis are satisfied, so if the residuals are normally distributed and independent of each other. It should be noted that, since the form of the histogram of the residuals strongly depends on the number of data and, as a consequence, on the number of used intervals, it is not always possible to obtain statistically significant information about the
	emerging distribution having less than 20 points. However, even if we had 16 data points, we have chosen to report this plot as well to understand if it would be possible to recover some information to be added to that obtained from the other residual plots.
	
	We report in the following the results of the statistical analysis for the two-qubit case $(N=2)$ regarding the estimation of the median and minimum {{PPTTs}}. For the two-qutrit case the same considerations hold and we will report, for the sake of completeness, a subsequent section with the tables related to the fit parameters' estimates in this last case.
	Let us now discuss the case $N=2$, all the quantities will be referred to. 
	{{\subsection{Estimation of the median PPTT $M_k$}}}
	The computed values of the parameters characterizing the fit function are the following
	\begin{table}[H]
		\centering
		\begin{tabular}
			{cccc}
			\toprule
			Param & Estimate & SE Estimate & 95\% CI \\
			\midrule
			$\theta_{1}$ & 1.0514 & 0.0007 & (1.0499; 1.0529) \\
			$\theta_{3}$ & 1.05 & 0.02 & (1.01; 1.09) \\		$\theta_{2}$ & 0.8599 & 0.0003 & (0.8592; 0.8605) \\
			\bottomrule
		\end{tabular}
		\caption{Parameter estimates for $M_k$ in the two-qubit case $(N=2)$.}
		\label{Tab:parameter_estimates_Me(tau_ent)}
	\end{table}
	The standard error of the estimate (SE Estimate) gives information about the precision of the computed parameters: the smaller its value, the more precise the estimate. We can notice that the error on parameter $\theta_{3}$ is about one or two orders of magnitude larger than the errors on parameters $\theta_{1}$ and $\theta_{2}$. In fact, while $\theta_{2}$ can be estimated precisely as it represents the value of the fit function when $\alpha$ goes to infinity, one expects a correlation between the other two parameters, which emerges in the correlation matrix \ref{Tab:Correlation_matrix_Me} reported below, that will reflect in a less precise estimation of these parameters, as they are both associated to the value of the fit function when $k=0$. In particular, from this consideration it becomes clear the choice we made on $\theta_{1}'$, which guarantees that just one of the latter two parameters has a larger error. The correlation matrix has the following expression
	\begin{table}[H]
		\centering
		\begin{tabular}{ccc}
			\toprule
			& $\theta_{1}$ & $\theta_{3}$\\
			\midrule
			$\theta_{3}$ & -0.590165 & \\
			$\theta_{2}$ & 0.087011 & -0.338910\\
			\bottomrule
		\end{tabular}
		\caption{Correlation matrix for $M_k$ in the two-qubit case $(N=2)$.}
		\label{Tab:Correlation_matrix_Me}
	\end{table}
	As expected, $\theta_{1}$ and $\theta_{3}$ are quite strongly correlated; this correlation could decrease if one considers more data points, however, given also the precision of the fit (see Fig.~\ref{fig:fit_t_ent_panel_a}), this has not been considered highly necessary. The residual plots are reported in Fig.~\ref{fig:t_ent_Me_residual_plots}.
	\begin{figure*}[!htb]
		\centering
		\includegraphics[scale=0.28]{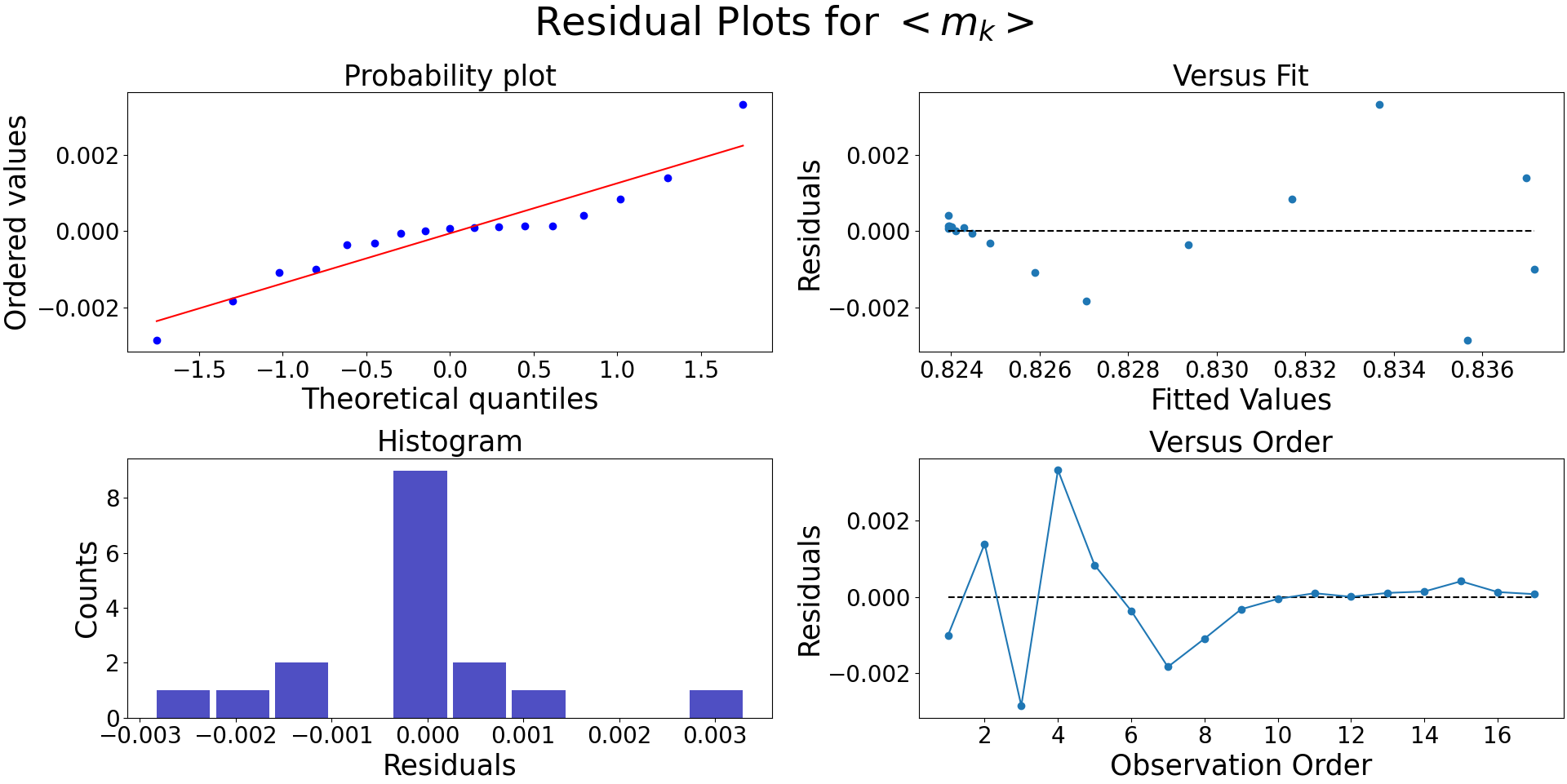}
		\caption{Residual plots for $m_k$ in the two-qubit case $(N=2)$ (maximum value of $k$ being $30$).}
		\label{fig:t_ent_min_residual_plots}
	\end{figure*}
	From these graphs it emerges that the residuals are of the order of the thousand and that there are some of them whose values are close to zero. These points are responsible for the variations from the expected pattern's behaviour, reported in the plot as a red line. In the normal probability plot, the residuals follow approximatively the red straight line and the presence of the aforementioned points is not so evident, while it emerges more clearly in the plot of the residuals versus fit and also in the plot of the residual versus order. In the former, one expects that the points are randomly distributed on both sides of the central dashed line. What can be observed is that there is a first group of residuals, that are randomly distributed in the range $[-10^{-2},\,10^{-2}]$, and a second group of them that, instead, have values in the range $[-3\cdot 10^{-2},\,3\cdot 10^{-2}]$.
	In particular, the smallest residuals are the ones obtained for larger $k$ values, as can be seen from the latter plot. However, when $k$ increases the function saturates to a certain value (given by $\theta_{2}$) and, as a consequence, the residuals tend to assume smaller values; there are not, in fact, sharp variations of the function as the ones that are present for smaller $k$ values or, in other words, the slope of the function becomes fixed. These considerations justify the obtained residual's graphs and permit us to conclude that the chosen fit function can be considered as a quite good estimate of the real function that links the considered time values to the Lindbladian parameters.
	{{
	\subsection{Estimation of the minimum PPTT $m_k$} 

		The previous analysis is repeated for the minimum value of the PPTT.}} As already pointed out in Sec.~\ref{subsec:analysis_char_EST}, one expects this time to be a less stable point of the distributions, since it strongly depends on the number of samplings that have been done. Indeed, this clearly appears in the plot Fig.~\ref{fig:fit_t_ent_panel_b}, where it can be seen that these points are characterized by larger errors than the previous case, especially for small $\alpha$ values. The parameter estimates are reported in the following table
		\begin{table}[H]
			\centering
			\begin{tabular}
				{cccc}
				\toprule
				Parameter & Estimate & SE Estimate & 95\% CI \\
				\midrule
				$\theta_{1}$ & 0.8372 & 0.0007 & (0.8358; 0.8386) \\
				$\theta_{3}$ & 0.7 & 0.2 & (0.4; 1.1) \\		$\theta_{2}$ & 0.8239 & 0.0003 & (0.8233; 0.8245) \\
				\bottomrule
			\end{tabular}
			\caption{Parameter estimates for $m_k$ in the two-qubit case $(N=2)$.}
			\label{Tab:parameter_estimates_min(tau_ent)}
			\end{table}
		Analogously to the previous cases, the best-estimated parameter is $\theta_{2}$. $\theta_{1}$ has a small error compared to $\theta_{3}$ which, instead, has a $20\%$ uncertainty that can be caused by the minor stability of this characteristic time. As previously, we report the correlation matrix of parameters
		\begin{table}[H]
			\centering
			\begin{tabular}{ccc}
				\toprule
				& $\theta_{1}$ & $\theta_{3}$\\
				\midrule
				$\theta_{3}$ & -0.591603 & \\
				$\theta_{2}$ & 0.080837 & -0.327771\\
				\bottomrule
			\end{tabular}
			\caption{Correlation matrix for parameter estimates for $m_k$ in the two-qubit case $(N=2)$.}
		\end{table}
		What emerges from this table is in line with what was found for the median time, therefore the same considerations hold. Finally, residual plots reported in Fig.~\ref{fig:t_ent_min_residual_plots} are shown to assess the goodness of fit. For these plots more careful considerations must be done. In these graphs it emerges the presence of a large number of points that have almost zero residuals. This is seen by the presence of points forming an almost horizontal line in the first graph while, instead, they should follow the red line, by the presence also of a higher concentration of points in the second graph, which therefore shows that the residuals have a different variance, and by the fact that, in the last graph, the points are no longer randomly distributed by increasing order, i.e.~by increasing $k$ values, which would imply, in principle, a correlation between adjacent points. As it can be seen from the last graph, these points correspond to the minimum times that one has for increasing $k$ values, when the histograms tend to a limit histogram and become more and more peaked in correspondence with the minimum time. In analyzing these graphs it is, therefore, necessary to take into account the behavior of the distributions that are being analyzed.
		Therefore, these residuals are not necessarily associated with an error in the model but could be conditioned by the trend of distributions.

\subsection{Parameters estimation of the median PPTT $M_k$ and the minimum PPTT $m_k$ for the two-qutrit case}
	For the sake of completeness, we report the values of parameter $\theta_{1}$, $\theta_{2}$, $\theta_{3}$ related to the fit function $f_{\vec{\theta}}(k)$ in Eq.~\ref{eq:fit_function}, and the correlation matrix of parameters, for the median and the minimum positive partial transpose time in the two-qutrit case, namely for $N=3$.
	
	Let us start from the median PPTT $M_{k}$. The fit parameters obtained from the statistical analysis are reported in the Table \ref{Tab:param_estim_M_k_N=3}, while the correlation matrix of parameter is given by Table \ref{Tab:correl_matr_m_k_N=3}.
	\begin{table}[H]
		\centering
		\begin{tabular}
			{cccc}
			\toprule
			Parameter & Estimate & SE Estimate & 95\% CI \\
			\midrule
			$\theta_{1}$ & 1.702 & 0.002 & (1.697;1.707) \\
			$\theta_{3}$ & 0.75 & 0.02 & (0.69;0.81) \\
			$\theta_{2}$ & 1.382 & 0.001 & (1.379;1.385)\\
			\bottomrule
		\end{tabular}
		\caption{Parameter estimates for $M_k$, in the two-qutrit case $(N=3)$.}
		\label{Tab:param_estim_M_k_N=3}
	\end{table}
	
	\begin{table}[H]
		\centering
		\begin{tabular}{ccc}
			\toprule
			& $\theta_{1}$ & $\theta_{3}$\\
			\midrule
			$\theta_{3}$ & -0.592114 & \\
			$\theta_{2}$ & 0.063677 & -0.335808\\
			\bottomrule
		\end{tabular}
		\caption{Correlation matrix for parameter estimates for $M_k$ in the two-qutrit case $(N=3)$.}
		\label{Tab:correl_matr_M_k_N=3}
	\end{table}
	
	For the minimum PPTT $m_{k}$, we report in Table \ref{Tab:param_estim_m_k_N=3} the parameter estimates and in Table \ref{Tab:correl_matr_m_k_N=3} the correlation matrix for the parameter estimates.
	
	\begin{table}[H]
		\centering
		\begin{tabular}
			{cccc}
			\toprule
			Parameter & Estimate & SE Estimate & 95\% CI \\
			\midrule
			$\theta_{1}$ & 1.383 & 0.003 & (1.377;1.390) \\
			$\theta_{3}$ & 0.59 & 0.05 & (0.48;0.73) \\
			$\theta_{2}$ & 1.242 & 0.002 & (1.238;1.245)\\
			\bottomrule
		\end{tabular}
		\caption{Parameter estimates for $m_k$, in the two-qutrit case $(N=3)$.}
		\label{Tab:param_estim_m_k_N=3}
	\end{table}
	
	\begin{table}[H]
		\centering
		\begin{tabular}{ccc}
			\toprule
			& $\theta_{1}$ & $\theta_{3}$\\
			\midrule
			$\theta_{3}$ & -0.586919 & \\
			$\theta_{2}$ & 0.065695 & -0.349262\\
			\bottomrule
		\end{tabular}
		\caption{Correlation matrix for parameter estimates for $m_k$ in the two-qutrit case $(N=3)$.}
		\label{Tab:correl_matr_m_k_N=3}
	\end{table}

	\section{Derivation of the Limit Lindbladian}{\label{app:Limit_Lindbladian}}
	We report here the explicit derivation of the limit Lindbladian starting from the expression of the Lindblad superoperator written in interaction picture: 
	\begin{widetext}
		\begin{equation}{\label{eq:Lindlad_ME_inter_pict}}
			\tilde{\mathcal{L}}_{H,K}^{(kt)}= \sum_{n=1}^{N^2 -1}\bigg[\tilde{L}_{n}^{(kt)}(\cdots)\tilde{L}_{n}^{{(kt)}\,\dagger} - \frac{1}{2}\bigg(\tilde{L}_{n}^{{(kt)}\,\dagger} \tilde{L}_{n}^{(kt)}(\cdots)\, +\, (\cdots) \tilde{L}_{n}^{{(kt)}\,\dagger}\tilde{L}_{n}^{(kt)}\bigg)\bigg],
		\end{equation}
	\end{widetext}
where:
\begin{equation}
	\tilde{L}_{n}^{(kt)} = e^{ik \hat{H} t} \hat{L}_K^{(n)} e^{-i k \hat{H} t}
\end{equation}
represents the Lindblad operator evolved in time.
As pointed out in Sec.~\ref{sect:LimLind}, an ansatz for the limit Lindbladian is obtained by considering the asymptotic average of   $\tilde{\cal L}_{K,H}^{(kt)}$ for large $k$, i.e.~:
	\begin{equation}
		\mathcal{L}_{H,K}^{(\text{eff})} \coloneqq \lim_{k \rightarrow \infty} \frac{1}{k} \int_{0}^{k}dk'\tilde{\mathcal{L}}_{H,K}^{(k't)}.
		\label{eq:limit_lindbladian_def_appendix}
	\end{equation}
	To perform the integral in Eq.~\eqref{eq:limit_lindbladian_def}, we divide it into three terms and compute them separately, namely
	\begin{gather}
		I_{1}^{(n)}(kt) = \frac{1}{k}\int_{0}^{k}\tilde{L}_{n}^{(k't)}(\cdots)\tilde{L}_{n}^{{(k't)}\,\dagger}\ dk' , \label{I1} \\
		I_{2}^{(n)}(kt) = \frac{1}{k}\int_{0}^{k}\tilde{L}_{n}^{{(k't)}\,\dagger}\tilde{L}_{n}^{(k't)}(\cdots)\ dk' , \label{I2} \\
		I_{3}^{(n)}(kt) = \frac{1}{k}\int_{0}^{k}(\cdots)\tilde{L}_{n}^{{(k't)}\,\dagger}\tilde{L}_{n}^{(k't)}\ dk' .
		\label{I3}
	\end{gather}
	The second and third integral will give the same result and one can compute only one of the two, namely, $I_{3}^{(n)}(t)$. This integral is performed only on Lindblad operators, which have the following expression
	
	\begin{equation}
		\tilde{L}_{n}^{{(k't)}\,\dagger}\tilde{L}_{n}^{(k't)} = e^{i k' \hat{H} t} \hat{L}_{K}^{(n)\,\dagger} \hat{L}_{K}^{(n)} e^{-i k' \hat{H} t}.
	\end{equation}
	It can be seen that $k'$ appears only in the exponential operator, which can be diagonalized because $\hat{H}$ is a Hermitian matrix, i.e.

	\begin{equation}
		e^{ik' \hat{H} t} = \sum_{l} e^{ik' \lambda_{l} t} \hat{\Pi}_{l}.
	\end{equation}
	In this way it is possible to separate the exponential coefficient, which contains $k'$ and the eigenvalues $\lambda_{l}$ of $\hat{H}$, from the projectors $\hat{\Pi}_{l}$ on the $l$-th eigenspace of the Hamiltonian. Therefore
	\begin{widetext}
	\begin{equation}
		\begin{aligned}
			\tilde{L}_{n}^{{(k't)}\,\dagger}\tilde{L}_{n}^{(k't)} & = \sum_{i,j} e^{ik' (\lambda_{i}-\lambda_{j}) t} \hat{\Pi}_{i} \hat{L}_{K}^{(n)\,\dagger} \hat{L}_{K}^{(n)} \hat{\Pi}_{j} = \sum_{i,j} e^{ik' t (\lambda_{i}-\lambda_{j})} \hat{W}_{i\,j}^{(n)}\\
			& = \sum_{i,j\neq i}e^{i k' t (\lambda_{i}-\lambda_{j})}\hat{W}_{i\,j}^{(n)} + \sum_{i,j}e^{ik' t (\lambda_{i}-\lambda_{j})}\hat{W}_{i\,j}^{(n)}\delta_{i,j}\\
			& = \sum_{i,j\neq i}e^{ik' t (\lambda_{i}-\lambda_{j})}\hat{W}_{i\,j}^{(n)} + \sum_{i}\hat{W}_{i\,i}^{(n)}.
		\end{aligned}
	\end{equation}
	\end{widetext}
	Once the integrand is explicitly stated in terms of $k'$, it is possible to compute the integral
		\begin{widetext}
			\begin{equation}
			\begin{split}
				I_{3}^{(n)}(kt) & = \frac{1}{k}\int_{0}^{k}(\cdots)\tilde{L}_{n}^{{(k't)}\,\dagger}\tilde{L}_{n}^{(k't)}\ dk'.\\
				& = (\cdots) \frac{1}{k}\int_{0}^{k}\sum_{i,j\neq i}e^{ik' t (\lambda_{i}-\lambda_{j})}\hat{W}_{i\,j}^{(n)}dk' + (\cdots) \frac{1}{k}\int_{0}^{k}\sum_{i}\hat{W}_{i\,i}^{(n)}dk'\\
				& = (\cdots) \sum_{i,j\neq i} \frac{i}{k t (\lambda_{j}-\lambda_{i})} \bigg( e^{ik t (\lambda_{i}-\lambda_{j})} -1 \bigg) \hat{W}_{i\,j}^{(n)} + (\cdots) \sum_{i} \hat{W}_{i\,i}^{(n)}.
			\end{split}
		\end{equation}
	\end{widetext}
	When $k \rightarrow \infty$ the first term goes to $0$ and $I_{3}^{(n)}(kt)$ becomes:
	\begin{equation}
		I_{3}^{(n)} = (\cdots) \sum_{i} \hat{W}_{i\,i}^{(n)}.
	\end{equation}
	The integral $I_{2}^{(n)}(kt)$ gives the same result as the integral $I_{3}^{(n)}(kt)$, with the only difference that the operator multiplies the matrix $\hat{W}_{i\,j}^{(n)}$ on the right. For this integral, the same considerations made for $I_{3}^{(n)}(kt)$ hold, therefore when $k \rightarrow \infty$, it becomes:
	\begin{equation}
		I_{2}^{(n)} = \sum_{i} \hat{W}_{i\,i}^{(n)} (\cdots).
	\end{equation}	
	\noindent Finally, it is possible to compute the integral $I_{1}^{(n)}(kt)$ (see Eq.~\eqref{I1}). As in the previous case, the dependence on $k$ in the integrand must be made explicit.
	Writing the Hamiltonian spectral decomposition, one has
	\begin{widetext}
	\begin{equation}
		\tilde{L}_{n}^{(k't)}(\cdots)\tilde{L}_{n}^{{(k't)}\,\dagger} =
		\sum_{i,j} \sum_{l,m} e^{ik't [(\lambda_{i} + \lambda_{l}) - (\lambda_{j}+\lambda_{m})]} \hat{\Pi_{i}} \hat{L}_{K}^{(n)} \hat{\Pi}_{j} (\cdots) \hat{\Pi}_{l} \hat{L}_{K}^{(n)\,\dagger} \hat{\Pi}_{m}.
		\label{L_theta_L_dagg}
	\end{equation}
	\end{widetext}
Defining the operator
\begin{eqnarray}
	\hat{Z}_{i,j}^{(n)}&:=&{\hat{\Pi}_i} {\hat{L}^{(n)}_K}{\hat{\Pi}_j}, \;
\end{eqnarray}
the expression in Eq.~\eqref{L_theta_L_dagg} can be written in a more compact form. Moreover, as in the previous case, the summation is split in two parts: a first part in which the exponential coefficient is different from one, and a second part in which the exponential coefficient is equal to one, so the dependence from $k'$ is lost.
	\begin{widetext}
	\begin{equation}
		\begin{split}
			\tilde{L}_{n}^{(k't)}(\cdots)\tilde{L}_{n}^{{(k't)}\,\dagger} & = \sum_{i,j} \sum_{l,m} e^{ik't [(\lambda_{i} + \lambda_{l}) - (\lambda_{j}+\lambda_{m})]} {\hat{Z}}_{i,j}^{(n)}(\cdots){\hat{Z}_{m,l}^{(n)\,\dagger}}\\
			& =\, \sum_{i,l}\sum_{\substack{j,m\, s.t. \\ (\lambda_{j}+\lambda_{m}) \neq (\lambda_{i}+\lambda_{l}) }} e^{ik' t [(\lambda_{i}+\lambda_{l})- (\lambda_{j}+\lambda_{m})]} \hat{Z}_{i,j}^{(n)}(\cdots){\hat{Z}_{m,l}^{(n)\,\dagger}}
			\quad \\
			& +\, \sum_{i,j} \sum_{l,m} e^{ik' t [(\lambda_{i}+\lambda_{l})- (\lambda_{j}+\lambda_{m})]}(\delta_{i,j}\delta_{l,m})\hat{Z}_{i,j}^{(n)} (\cdots){\hat{Z}_{m,l}^{(n)\,\dagger}} \quad\\
			& +\, \sum_{i,j} \sum_{m,l\neq i} e^{ik' t [(\lambda_{i}+\lambda_{l})- (\lambda_{j}+\lambda_{m})]}(\delta_{j,l} \delta_{i,m})\hat{Z}_{i,j}^{(n)} (\cdots){\hat{Z}_{m,l}^{(n)\,\dagger}}\ \\
			& =\, \sum_{i,l}\sum_{\substack{j,m\, s.t. \\ (\lambda_{j}+\lambda_{m}) \neq (\lambda_{i}+\lambda_{l}) }} e^{ik' t [(\lambda_{i}+\lambda_{l})- (\lambda_{j}+\lambda_{m})]} \hat{Z}_{i,j}^{(n)}(\cdots){\hat{Z}_{m,l}^{(n)\,\dagger}}
			\quad \\
			& +\, \sum_{i,l}\hat{Z}_{i,i}^{(n)}(\cdots) \hat{Z}_{l,l}^{(n)\,\dagger} + \sum_{i,l\neq i} \hat{Z}_{i,l}^{(n)}(\cdots) \hat{Z}_{i,l}^{(n)\,\dagger}.
		\end{split}
	\end{equation}
\end{widetext}
	In this way, the first term depends on $k'$ while the second term does not. Therefore, when integrating on $k'$, the only non trivial integral will be the first one. The form of the integrand function is the same as the previous case; as a consequence, the resulting integral will be a function of the eigenvalues of $\hat{H}$ and will depend on $k$ in such a way that, when $k \rightarrow \infty$, it becomes negligible. Calling $g(k,t,\lambda_{i},\lambda_{j},\lambda_{l},\lambda_{m})$ the resulting integral function, the final form of $I_{1}^{(n)}(kt)$ is
	\begin{widetext}
		\begin{equation}
		\begin{split}
			I_{1}^{(n)}(kt) & = \sum_{i,l}\sum_{\substack{j,m\, s.t. \\ (\lambda_{j}+\lambda_{m}) \neq (\lambda_{i}+\lambda_{l}) }} g(k,t,\lambda_{i},\lambda_{j},\lambda_{l},\lambda_{m}) \hat{Z}_{i,j}^{(n)}(\cdots) \hat{Z}_{m,l}^{(n)\,\dagger} \quad \\
			& \, + \sum_{i,l}\hat{Z}_{i,i}^{(n)}(\cdots) \hat{Z}_{l,l}^{(n)\,\dagger} + \sum_{i,l\neq i} \hat{Z}_{i,l}^{(n)}(\cdots) \hat{Z}_{i,l}^{(n)\,\dagger}.
		\end{split}
		\end{equation}
	\end{widetext}
	When $k \rightarrow \infty $, the function $g(k,t,\lambda_{i},\lambda_{j},\lambda_{l},\lambda_{m}) \rightarrow 0$ and $I_{1}^{(n)}(kt)$ becomes:
	\begin{equation}
		I_{1}^{(n)} = \sum_{i,l}\hat{Z}_{i,i}^{(n)}(\cdots) \hat{Z}_{l,l}^{(n)\,\dagger} + \sum_{i,l\neq i} \hat{Z}_{i,l}^{(n)}(\cdots) \hat{Z}_{i,l}^{(n)\,\dagger}.
	\end{equation}	
	Putting together the terms $I_{1}^{(n)}, \ I_{2}^{(n)}, \ I_{3}^{(n)}$, one obtains the final expression of the limit Lindbladian, when $k \rightarrow \infty$
	\begin{widetext}
		\begin{equation}
		\mathcal{L}_{H,K}^{(\text{eff})} = \sum_{n = 1}^{N^2 - 1} \Bigg[ \sum_{i,l}\hat{Z}_{i,i}^{(n)}(\cdots) \hat{Z}_{l,l}^{(n)\,\dagger} + \sum_{i,l\neq i} \hat{Z}_{i,l}^{(n)}(\cdots) \hat{Z}_{i,l}^{(n)\,\dagger} - \frac{1}{2} \Big( \sum_{i} \hat{W}_{i\,i}^{(n)} (\cdots)  + (\cdots)  \sum_{i} \hat{W}_{i\,i}^{(n)} \Big) \Bigg].
		\end{equation}
	\end{widetext}
	The operator $W_{i\,i}^{(n)}$ can be written in terms of $Z^{(n)}$ operators, so $\mathcal{L}_{H,K}^{(\text{eff})}$ assumes the following equivalent expression
	\begin{widetext}
	\begin{equation}
		\begin{split}
			\mathcal{L}_{H,K}^{(\text{eff})} = \sum_{n = 1}^{N^2 - 1} \Bigg[ \sum_{i,l}\hat{Z}_{i,i}^{(n)}(\cdots) \hat{Z}_{l,l}^{(n)\,\dagger} + \sum_{i,l\neq i} \hat{Z}_{i,l}^{(n)}(\cdots) \hat{Z}_{i,l}^{(n)\,\dagger} - \frac{1}{2} \sum_{i, l} \Big( \hat{Z}_{l,i}^{(n)\,\dagger}  \hat{Z}_{l,i}^{(n)}(\cdots) + (\cdots) \hat{Z}_{l,i}^{(n)\,\dagger} \hat{Z}_{l,i}^{(n)} \Big) \Bigg].
			\label{eq:L_lim_app}
		\end{split}
	\end{equation}
	\end{widetext}
	
	To specify this derivation for example in the two-qubit case, where $N=2$, the expression of the limit Lindbladian \eqref{eq:L_lim_Z} (here Eq.~\eqref{eq:L_lim_app}) reduces to \eqref{eq:L_lim_with_A} considering the index conversion $(i,l)\rightarrow m = 2i + l$, with $i=0,1$ and $l=0,1$, and introducing the following symmetric matrix $A$
	
	\begin{equation}
		A = 
		\begin{pmatrix}
			1 & 0 & 0 & 1\\
			0 & 1 & 0 & 0\\
			0 & 0 & 1 & 0\\
			1 & 0 & 0 & 1
		\end{pmatrix}.
	\end{equation}
	By diagonalizing it, one can obtain the Lindblad matrices, defined in Eq.~\eqref{eq:Lim_lind_Lindbald_matrices}, that enters in the final expression of the limit Lindbladian \eqref{eq:L_lim_lindblad_form}
	\begin{equation}
		\label{eqn:schema}
		\begin{split}
			& \hat{Y}^{(0,n)} = 0, \\
			& \hat{Y}^{(1,n)} = \hat{Z}_{2}^{(n)},\\
			& \hat{Y}^{(2,n)} = \hat{Z}_{1}^{(n)},\\
			& \hat{Y}^{(3,n)} = -(\hat{Z}_{0}^{(n)} + \hat{Z}_{3}^{(n)}).
		\end{split}
	\end{equation}

\end{document}